\documentclass[useAMS,usenatbib]{mn2e}
\usepackage{color}
\usepackage{graphicx}
\usepackage{graphics}
\usepackage{rotating}
\usepackage{amsmath}        
\usepackage{amssymb}        
\usepackage{subfigure}        
\usepackage{epsfig}        
\usepackage{longtable}
\usepackage[T1]{fontenc}
\usepackage{aecompl}

\newcommand{\mnras}{MNRAS}
\newcommand{\apjl}{ApJ}
\newcommand{\apj}{ApJ}
\newcommand{\apjs}{ApJS}
\newcommand{\aj}{AJ}
\newcommand{\nat}{Nature}
\newcommand{\aap}{A\&A}

\newcommand{\pasa}{PASA}
\newcommand{\apss}{Ap\&SS}
\newcommand{\araa}{ARA\&A}
\newcommand{\pasp}{PASP}
\newcommand{\na}{NA}
\newcommand{\bain}{BAIN}

\newcommand{\source}{MAXI\,J1836$-$194}

\newcommand{\degree}{$^{\circ}$}

\newcommand{\arcsecond}{$^{\prime\prime}$}
\newcommand{\perbeam}{\,beam$^{-1}$}

\title[Radio monitoring of the hard state jets in \source]{Radio monitoring of the hard state jets in the 2011 outburst of \source{}}

\author[T.~D.~Russell et al.]
 {T.~D.~Russell,$^{1}$\thanks{email: thomas.russell@icrar.org} J.~C.~A.~Miller-Jones,$^{1}$ P.~A.~Curran,$^{1}$ R.~Soria,$^{1}$ D.~Altamirano,$^{2}$ \and S.~Corbel,$^{3,4}$ M. Coriat,$^{5}$ A.~Moin,$^{6,7}$ D.~M.~Russell,$^{6}$ G.~R.~Sivakoff,$^{8}$ \and T.~J.~Slaven-Blair,$^{1}$ T.~M.~Belloni,$^{9}$ R.~P.~Fender,$^{10}$  S.~Heinz,$^{11}$ P. G. Jonker,{$^{12,13,14}$} \and H.~A.~Krimm,$^{15,16}$ E.~G.~K\"ording,$^{14}$ D.~Maitra,$^{17}$ S.~Markoff,$^{18}$ M.~Middleton,$^{19}$ \and S.~Migliari,$^{20}$ R.~A.~Remillard,$^{21}$ M.~P.~Rupen,$^{22,23}$ C.~L.~Sarazin,$^{24}$ A.~J.~Tetarenko,$^{8}$ \and  M. A. P. Torres,$^{12,14}$ V.~Tudose,$^{25}$ and  A. K. Tzioumis,$^{26}$\\
$^1$International Centre for Radio Astronomy Research - Curtin University, GPO Box U1987, Perth, WA 6845, Australia\\
$^2$School of Physics and Astronomy, University of Southampton, Highfield SO17 IBJ, England\\
$^3$Laboratoire AIM (CEA/IRFU - CNRS/INSU - Universit\'{e} Paris Diderot), CEA DSM/IRFU/SAp, F-91191 Gif-sur-Yvette, France\\
$^{4}$Station de Radioastronomie de Nan\c{c}ay, Observatoire de Paris, CNRS/INSU, USR 704 - Univ. Orl\'{e}ans, OSUC, 18330 Nan\c{c}ay, France\\
$^5$Department of Astronomy, University of Cape Town, Private Bag X3, Rondebosch 7701, South Africa\\
$^6$New York University Abu Dhabi, P.O. Box 129188, Abu Dhabi, United Arab Emirates \\
$^{7}$Shanghai Astronomical Observatory, 80 Nandan Road, Xujiahui, Shanghai 200030, China\\
$^{8}$Department of Physics, University of Alberta, 4-181 CCIS, Edmonton, AB T6G 2E1, Canada\\
$^{9}$INAF - Osservatorio Astronomico di Brera, Via E. Bianchi 46, I-23807, Merate (LC), Italy \\
$^{10}$Department of Physics, Oxford University, Denys Wilkinson Building, Keble Road, Oxford OX1 3RH, UK \\
$^{11}$Astronomy Department, University of Wisconsin-Madison, 475. N. Charter St., Madison, WI 53706, USA \\
$^{12}$SRON, Netherlands Institute for Space Research, Sorbonnelaan 2, NL-3584 CA Utrecht, the Netherlands \\
$^{13}$Harvard-Smithsonian Center for Astrophysics, 60 Garden Street, Cambridge, MA 02138, USA \\
$^{14}$Department of Astrophysics/IMAPP, Radboud University Nijmegen, PO Box 9010, NL-6500 GL Nijmegen, the Netherlands \\
$^{15}$NASA/Goddard Space Flight Center, Greenbelt, MD 20771, USA\\ 
$^{16}$USRA, 10211 Wincopin Circle, Suite 500, Columbia, MD 21044, USA\\
$^{17}$Department of Physics and Astronomy, Wheaton College, Norton, MA 02766, USA \\
$^{18}$Astronomical Institute `Anton Pannekoek', University of Amsterdam, P.O. Box 94249, 1090 GE Amsterdam, the Netherlands \\
$^{19}$Institute of Astronomy, Madingley Rd, Cambridge CB3 0HA, UK\\
$^{20}$Department of Astronomy and Meteorology \& Institute of Cosmic Sciences, University of Barcelona, Mart\'{i} i Franqu\`{e}s 1, \\ 08028 Barcelona, Spain. \\
$^{21}$MIT Kavli Institute for Astrophysics and Space Research, Building 37, 70 Vassar Street, Cambridge, MA 02139, USA \\
$^{22}$National Research Council, Herzberg Astronomy and Astrophysics, 717 White Lake Road, PO Box 248, Penticton, \\British Columbia V2A 6J9, Canada\\
$^{23}$National Radio Astronomy Observatory, P.O. Box 0, Socorro, NM 87801, USA\\
$^{24}$Department of Astronomy, University of Virginia, P.O. Box 400325, Charlottesville, VA 22904, USA \\
$^{25}$Institute for Space Sciences, Atomistilor 409, P.O. Box MG-23, Bucharest-M\v{a}gurele RO-077125, Romania \\
$^{26}$CSIRO Astronomy and Space Science, ATNF, PO Box 76, Epping, NSW, 1710, Australia\\
}
\begin{document}
\date{Accepted 2015 March 30.}
\pagerange{\pageref{firstpage}--\pageref{lastpage}} \pubyear{2015}
\maketitle

\label{firstpage}
\begin{abstract}
\source{} is a Galactic black hole candidate X-ray binary that was discovered in 2011 when it went into outburst. In this paper, we present the full radio monitoring of this system during its `failed' outburst, in which the source did not complete a full set of state changes, only transitioning as far as the hard intermediate state. Observations with the Karl G. Jansky Very Large Array (VLA) and Australia Telescope Compact Array (ATCA) show that the jet properties changed significantly during the outburst. The VLA observations detected linearly polarised emission at a level of $\sim$1\% early in the outburst, increasing to $\sim$3\% as the outburst peaked. High-resolution images with the Very Long Baseline Array (VLBA) show a $\sim$15\,mas jet along the position angle $-21\pm2^\circ$, in agreement with the electric vector position angle found from our polarisation results ($-21\pm4^\circ$), implying that the magnetic field is perpendicular to the jet. Astrometric observations suggest that the system required an asymmetric natal kick to explain its observed space velocity. Comparing quasi-simultaneous X-ray monitoring with the 5\,GHz VLA observations from the 2011 outburst shows an unusually steep hard-state radio/X-ray correlation of $L_{\rm R} \propto L_{\rm X}^{1.8\pm0.2}$, where $L_{\rm R}$ and $L_{\rm X}$ denote the radio and X-ray luminosities, respectively. With ATCA and {\it Swift} monitoring of the source during a period of re-brightening in 2012, we show that the system lay on the same steep correlation. Due to the low inclination of this system, we then investigate the possibility that the observed correlation may have been steepened by variable Doppler boosting.

\end{abstract}


\section{Introduction}

Relativistic jets are observed from actively-accreting black holes on all scales, from stellar-mass black holes in low-mass X-ray binary (LMXB) systems to supermassive black holes at the centres of galaxies. There is a clear but poorly understood relationship between the morphology and spectrum of the jets and the structure of the accretion flow \citep[e.g.][]{2010LNP...794..115F,2010LNP...794...85G}. LMXBs provide ideal laboratories to study jets and their connection to the accretion flow as they evolve through their full duty cycles on humanly-observable timescales. Spending the majority of their lifetimes in a low-luminosity quiescent state, LMXBs occasionally go through a much brighter outburst phase (the result of increased accretion onto the central compact object) where the jet properties change significantly as the system transitions through a number of canonical X-ray states (see \citealt{2011BASI...39..409B} for a full review of X-ray spectral states).

In a typical outburst, LMXBs are initially observed in the hard state (HS), which is associated with persistent radio emission from a steady, partially self-absorbed, compact jet \citep{2000ApJ...543..373D,2000A&A...359..251C, 2001MNRAS.327.1273S} that has a flat or inverted, optically-thick radio spectrum ($\alpha \ga 0$, where $S_{\rm \nu} \propto \nu^{\alpha}$; \citealt{2001MNRAS.322...31F}). As the outburst progresses, the accretion rate increases and the system may transit through a number of intermediate states as it approaches the soft state. During the transition to the full soft state the jet emission transforms into discrete, bright, relativistically-moving knots with an optically-thin radio spectrum \citep{2004ApJ...617.1272C,2004MNRAS.355.1105F}. The compact jet is quenched by at least 2.5 orders of magnitude \citep{1999ApJ...519L.165F,2011MNRAS.414..677C,2011ApJ...739L..19R} as the jet spectral break moves to lower frequencies (\citealt{2013MNRAS.428.2500C,2013ApJ...768L..35R}; \citealt{2014MNRAS.439.1390R}) and particle acceleration seemingly occurs at different locations within the jet \citep{2014MNRAS.439.1390R}. At the end of the outburst, the X-ray luminosity decreases and the source transitions back to the hard state. In this reverse state transition the compact jet is gradually re-established (first switching on in the radio band and then the infrared/optical band; \citealt{2012MNRAS.421..468M,2013MNRAS.428.2500C,2013ApJ...779...95K}), indicating the potential build up of particle accelerating structures increasingly closer to the black hole.

The processes that are responsible for launching the jet outflow are poorly understood. However, there exists a well-studied relationship between the radio and X-ray luminosities in hard state LMXBs, which has been observed for both individual sources (e.g. GX 339$-$4; \citealt{1998A&A...337..460H,2003A&A...400.1007C,2013MNRAS.428.2500C}) and the entire sample of BH X-ray binaries (e.g. \citealt{2003MNRAS.344...60G}). Although recent work has cast some doubt over the universality of the relation \citep{2014MNRAS.445..290G}, it is generally described by two distinct power-law tracks in the radio/X-ray luminosity plane, an upper track and a lower, steeper, `radio-quiet' track \citep{2012MNRAS.423..590G,2011MNRAS.414..677C,2013MNRAS.428.2500C}. The track followed by a given source does not appear to be connected to any property of the object (such as spin; \citealt{2011MNRAS.413.2269S}) and there is evidence for some individual sources transitioning between the two tracks as their X-ray luminosity fades \citep{2011MNRAS.414..677C,2012MNRAS.423.3308J,2012MNRAS.423.2656R}. It has also been suggested that this bimodal relationship may also exist in AGN, supporting the theory that the mechanisms responsible for extracting energy from the accretion flow are similar across the mass scale \citep{2013ApJ...774L..25K}.

High resolution, very long baseline interferometry (VLBI) observations can reveal important information about the changing properties of the jet during an outburst. Directly imaging the jet can determine the orientation of the jet axis, the bulk motion and structure of the jet, and how these change with the evolving accretion flow. While discrete ejecta have been observed in a number of systems in outburst \citep[e.g.][]{1994Natur.371...46M,1995Natur.374..141T,1995Natur.375..464H,1999MNRAS.304..865F}, the hard state compact jet has only been spatially resolved in two systems, GRS 1915$+$105 \citep{2000ApJ...543..373D} and Cygnus X$-$1 \citep{2001MNRAS.327.1273S}, both of which are not standard LMXBs. High-cadence observations of the discrete ejecta can determine their proper motions. Extrapolating backwards in time, the moment of launching may then be coupled to changes within the accretion flow to identify the mechanisms responsible for launching \citep{2012MNRAS.421..468M}. Astrometric observations of a system over time may also allow for the proper motion of the system to be calculated, which can then be used to constrain the formation mechanism of black holes \citep[for full discussion, see][and references therein]{2014PASA...31...16M}. 

\subsection{\source{}}
\source{} was discovered at the onset of its outburst in August 2011, triggering an intensive multiwavelength observing campaign. Classified as a black hole candidate due to its radio and X-ray properties \citep{2011ATel.3618....1S,2011ATel.3628....1M,2011ATel.3689....1R,2012ApJ...751...34R, 2014MNRAS.439.1381R}, \source{} transitioned from the hard state \citep[spectral states are defined by X-ray spectral and timing properties;][]{ferrigno2012} to the hard intermediate state (HIMS) on 2011 September 11. The outburst `failed' to enter the full soft state, remaining in the HIMS (see \citealt{2004NewA....9..249B} for further discussion on failed outbursts), reaching its softest point on September 16, following which it transitioned back to the hard state on September 28 and faded towards quiescence. In 2012 March, the system underwent a period of renewed activity \citep{2012ATel.3966....1K,2012ATel.3975....1Y,2013AstL...39..367G} before returning to quiescence by 2012 July 05 \citep{2012ATel.4255....1Y}. This low-inclination system (between 4$^{\circ}$ and 15$^{\circ}$; \citealt{2014MNRAS.439.1381R}) is between 4 and 10 kpc away \citep{2014MNRAS.439.1390R} and has proven itself to be an ideal system in which to study the evolution of the radio jet due to its bright radio emission (\citealt{2013ApJ...768L..35R}; \citealt{2014MNRAS.439.1390R}). In this paper we present results from an intensive VLA, ATCA and VLBA radio monitoring campaign of \source{}, which began during the early phase of the outburst (2011 September 3) and continued until the source became sun constrained (our last radio observation occurred on 2011 December 3). We also present ATCA and quasi-simultaneous {\it Swift} monitoring of the source during its 2012 re-flare. In particular, we will discuss the evolution of the polarised radio jet, the jet morphology, and the VLBI astrometry during the 2011 outburst, as well as the hard state radio/X-ray correlation as the system faded towards quiescence.


\section{Observations}
\label{sec:observations}

\subsection{Radio observations}

\subsubsection{VLA}
\label{sec:VLA}

From 2011 September 3 through December 3, we conducted an intensive monitoring campaign on \source{} with the Karl G. Jansky Very Large Array (VLA), under project code VLA/10B-218. We have previously presented the six epochs of VLA observations with simultaneous millimetre, mid-/near-infrared, optical and X-ray data (\citealt{2013ApJ...768L..35R}; \citealt{2014MNRAS.439.1390R}). Here, we report the full 18 epochs of VLA radio observations covering the rise, transition to and from the HIMS, and subsequent decay of the 2011 outburst of \source{}. The VLA observations were taken every 2--5 days during the brighter phase of the outburst (September 3 -- September 30) and every 1 -- 2 weeks as the outburst decayed, until the final observation on 2011 December 3. The array was in its most-extended A configuration until September 12, in the A$\rightarrow$D move-configuration until September 18, and thereafter in the most-compact D configuration.

The observations were taken over a wide range of frequencies (between 1 and 43\,GHz) with an integration time of 3\,seconds for all observations prior to  November 11, following which the data were taken with an integration time of 5 seconds. The lowest-frequency 1--2\,GHz observations were recorded in two 512-MHz basebands, each containing eight 64-MHz sub-bands made up of sixty-four 1-MHz channels. All other frequencies were recorded in two 1024-MHz basebands comprised of eight 128-MHz sub-bands made up of sixty-four 2-MHz channels. Following standard data reduction procedures within the Common Astronomy Software Application (CASA; \citealt{2007ASPC..376..127M}) the data were corrected for shadowing, instrumental issues, and radio frequency interference. 3C286 was used for bandpass and amplitude calibration, J1407$+$2827 was used as a polarisation leakage calibrator, and phase calibration was performed using the nearby ($\sim 7^{\circ}$ away) compact source J1820$-$2528. Calibrated data were then imaged on a per-baseband basis using the multi-frequency synthesis algorithm within CASA and subjected to multiple rounds of self-calibration. To determine the flux density of the source, we fitted a point source to the target in the image plane (Stokes $I$) and standard VLA systematic errors were added. Stokes $Q$ and $U$ were also measured at the position of the peak source flux in the lower-frequency ($<$10\,GHz) data (Table~\ref{tab:VLA_flux}). At no epoch was the source observed to be significantly extended. 

\begin{figure}
\centering
\includegraphics[width=\columnwidth]{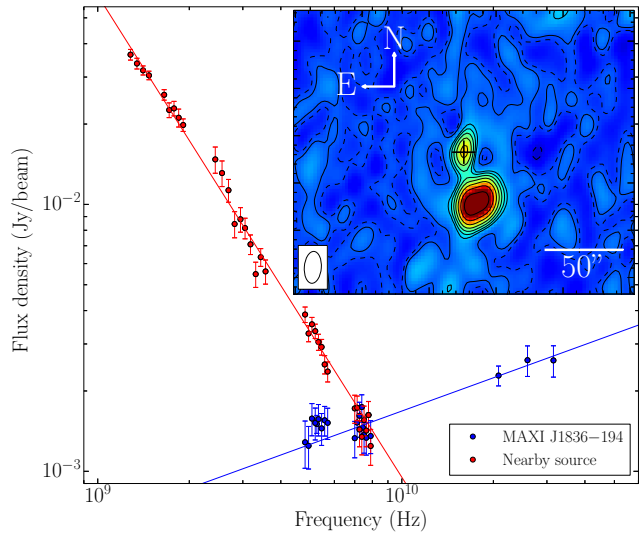}
\caption{VLA spectrum of \source{} (blue points) and the nearby unrelated source (red points) from 2011 November 1. At frequencies where the nearby object could not be resolved from our target ($<$4\,GHz) we subtracted the expected \source{} brightness from the combined flux densities to approximate the emission from the unrelated source. Inset: November 1 VLA 5\,GHz image of \source{} (black cross), showing the confusing source to the south west of the target. Contour levels are at $\pm(\sqrt{2}) ^n$ times the rms noise (0.1~mJy), where $n=-3,3,4,5 \dots$ (dashed contours represent negative values). At low frequencies ($<$4\,GHz) when the VLA was in the compact D configuration, \source{} was unresolved from the nearby unrelated object. Fitting $S_{\nu}$ against $\nu$ with a power law, we then estimate the spectral index of the confusing source to be $-$1.81$\pm$0.02. }
\label{fig:source}
\end{figure}

A nearby source is located to the south-west of \source{} ($\sim$30\arcsecond away; see Figure~\ref{fig:source}, inset). This unrelated, steep-spectrum source was not aligned with the jet axis and was not variable over the duration of the outburst. At times when the VLA was in its most compact D-configuration (for all observations after 2011 September 18), our low-frequency ($<$4\,GHz) VLA observations could not resolve the target from the confusing source. Therefore, where unresolved, our observations of \source{} are contaminated by the nearby source. We must determine the spectrum of the unrelated object to estimate its contribution and then subtract from necessary observations. Using an epoch when \source{} was faint (2011 November 01), we measured the brightness of the confusing source (Figure~\ref{fig:source}). At frequencies where we were unable to resolve the two sources from each other, we subtracted the expected \source{} brightness (determined by fitting a power law to the higher frequencies) from the total spectrum to give an approximation of the unrelated source. Fitting the measured radio flux densities, $S_{\nu}$, against frequency, $\nu$, with a power law, we then estimate the spectrum of the nearby source and \source{}. The spectrum of \source{} was well fit by a single power law at all epochs.

\begin{table*}
\caption{Sample VLA flux densities of \source{}. $\alpha$ is the radio spectral index. 1$\sigma$ errors are uncertainties on the fitted source parameters. Stokes $Q$ and $U$ are given before systematic errors are added. The full table is available online. }
\centering
\label{tab:VLA_flux}
\begin{tabular}{ccccccc}
\hline
Date & MJD & Frequency & Flux density & $Q$ & $U$ & $\alpha$ \\
 (UT) & & (GHz) & (mJy) &(mJy beam$^{-1}$)& (mJy beam$^{-1}$) &  \\
\hline
2012 Sep 03 & 55807.1 & 4.60  & 27.2$\pm$0.3& 0.24$\pm$0.05 & -0.29$\pm$0.06& 0.64$\pm$0.03 \\
 &  & 7.90  & 38.4$\pm$0.5& 0.67$\pm$0.06 & -0.49$\pm$0.07& \\
2011 Sep 05 & 55809.1 & 5.00  & 23.2$\pm$0.3& 0.08$\pm$0.02 & -0.08$\pm$0.02& 0.84$\pm$0.04\\
 &  & 7.45  & 32.5$\pm$0.4& 0.18$\pm$0.02 & -0.16$\pm$0.02& \\
2011 Sep 07 & 55811.2 & 1.50  & 18.1$\pm$0.4& -- & --& 0.51$\pm$0.01 \\
& & 5.00  & 28.6$\pm$0.7& 0.25$\pm$0.07 & -0.13$\pm$0.07& \\
& & 7.45  & 39.3$\pm$0.9& 0.58$\pm$0.09 & -0.08$\pm$0.08& \\
& & 20.80  & 68$\pm$2& -- & --& \\
& & 25.90  & 79$\pm$2& -- & --& \\
& & 32.02  & 88$\pm$5& -- & --& \\
\hline
\end{tabular}
\end{table*}

\subsubsection{VLBA}
\label{sec:VLBA}

\begin{table*}
\caption{Elliptical Gaussian fits (in the {\it uv}-plane) to the 2.3 and 8.4\,GHz VLBA images of MAXI J1836$-$194, showing the flux density, beam size, major axis, minor axis and the position angle for each VLBA epoch. The radio flux brightened during the X-ray peak of the outburst and the major axis size remained steady at 8.4\,GHz during our observations. }
\centering
\label{tab:xband_fits}
\begin{tabular}{cccccccc}
\hline
Date & MJD & Freq.& Flux & Beam & Major & Minor & Position\\
2011 &  & (GHz) & density & size & axis & axis & Angle\\
(UT)&  &       &  (mJy) & (mas) & ($\mu$as) & ($\mu$as) & (\degree)\\
\hline
Sep 04 & $55808.12\pm0.06$ & 8.42 & $30.8\pm0.2$ & $2.5\times0.8$ & $584\pm52$ & $188\pm27$ & $-17.0\pm4.6$\\
& & 2.27 & $15.9\pm0.3$ & $16.5\times4.8$ & $2950\pm200$ & $2180\pm1380$ & $90.0\pm0.1$\\
Sep 09 & $55813.08\pm0.08$ & 8.42 & $38.5\pm0.2$ & $2.6\times0.8$ & $699\pm33$ & $140\pm45$ & $-25.7\pm3.3$\\
& & 2.27 & $24.6\pm0.3$ & $12.2\times3.4$ & $6810\pm320$ & $2180\pm170$ & $-17.5\pm2.5$\\
Sep 19 & $55823.06\pm0.08$ & 8.42 & $29.3\pm0.1$ & $2.4\times0.8$ & $685\pm32$ & $101\pm46$ & $-25.6\pm3.0$\\
& & 2.27 & $28.9\pm0.2$ & $14.3\times4.4$ & $4870\pm310$ & $2830\pm110$ & $-18.9\pm5.7$\\
Oct 12 & $55846.02\pm0.10$ & 8.42 & $5.0\pm0.1$ & $2.3\times0.8$ & $833\pm123$ & $0\pm395$ & $-11.8\pm4.5$\\
& & 2.27 & $3.4\pm0.2$ & $19.6\times0.8$ & $9850\pm1750$ & $2\pm2730$ & $-15.5\pm5.5$\\
Nov 18 & $55883.88\pm0.10$ & 8.42 & $0.36\pm0.09$ & $2.8\times1.0$ & $-$ & $-$ & $-$ \\
\hline
\end{tabular}
\end{table*}

\source{} was observed five times with the VLBA during the outburst, for between 4 and 6\,hours per epoch, under project code BM339 (Table~\ref{tab:xband_fits}). We observed in dual polarisation mode, with 64\,MHz of bandwidth per polarisation.  In the first four epochs, we made use of the dual 13/4-cm recording mode to split the bandwidth equally between two frequency bands centred at 8.4 and 2.25\,GHz.  However, since our VLA monitoring indicated the source flux density to be $<1$\,mJy at the time of the final epoch, we observed at 8.4\,GHz only on November 18, using the full 64\,MHz of available bandwidth.

The observations were phase-referenced, using the closest available extragalactic calibrator source, J1832-2039, from the VLBA Calibrator Survey \citep[VCS;][]{2002ApJS..141...13B}, whose assumed position (from the most up-to-date global astrometric solution available at the time of the observations\footnote{http://astrogeo.org/vlbi/solutions/rfc\_2011c/}) was $18^{\rm h}32^{\rm m}11^{\rm s}.0465$, $-20^{\circ}39^{\prime}48^{\prime\prime}.202$ (1.6$^\circ$ from the target source). The phase referencing cycle time was 3\,minutes, spending 2\,minutes on the target source and 1\,minute on the phase reference calibrator in each cycle.  Every seventh observation of the target was substituted for a 1-minute observation of an astrometric check source.  We used two different check sources, alternating between J1845-2200, taken from the Fourth VLBA Calibrator Survey \citep[VCS-4;][]{2006AJ....131.1872P}, and J1825-1718, from the Second VLBA Calibrator Survey \citep[VCS-2;][]{2003AJ....126.2562F}.  30\,minutes at the start and end of each observation were used for geodetic blocks, observing several bright extragalactic calibrators over a range of elevations.  This improved the accuracy of the phase referencing process by allowing us to solve for unmodelled tropospheric delays and clock errors.

The data were correlated using the VLBA-DiFX implementation of the software correlator developed by \citet{2007PASP..119..318D}, and reduced using the Astronomical Image Processing System \citep[{\sc aips};][]{2003ASSL..285..109G}, according to standard procedures.  Data with elevations $<15^\circ$ were flagged, to prevent the increased atmospheric phase fluctuations at low elevations from affecting the astrometric accuracy.  Scatter-broadening affected the data at 2.25\,GHz, reducing the amplitude measured on the longer baselines, and we were unable to calibrate the station at Mauna Kea (MK).  Phase solutions for MAXI J1836-194 were derived by fringe-fitting the phase reference calibrator, using as a model the best available image of the calibrator following self-calibration.

For the first four epochs, the data were imaged with robust weighting as a compromise between sensitivity and resolution.  For the first three epochs, phase-only self-calibration was performed down to a solution interval of 5\,minutes.  These phase corrections were applied, before a single round of amplitude and phase calibration was used to adjust the overall amplitude gains of the individual antennas. By the fourth epoch, the source was significantly fainter, and we were only able to perform a single round of phase-only self calibration on a timescale of 1 hour. For the faint, fifth epoch, we imaged the data with natural weighting to maximise the sensitivity, but the source was too faint for self-calibration.

Since the final images showed the source to be marginally resolved in the first four epochs, we exported the self-calibrated data to \textsc{Difmap} \citep{1997ASPC..125...77S}, and fit the source with a single elliptical Gaussian in the uv-plane.  Since \textsc{Difmap} does not provide uncertainties on fitted model parameters, we used the best-fitting values as initial estimates for the \textsc{AIPS} task \textsc{uvfit}, which was used to determine the final fitted source sizes.

\subsubsection{ATCA}
\label{sec:ATCA}

The Australia Telescope Compact Array (ATCA) observed \source{} four times during its 2011 outburst and five times in 2012 (Table~\ref{tab:ATCA_flux}). All observations were taken at 5.5 and 9.0\,GHz with a bandwidth of 2\,GHz at each frequency. The ATCA data were reduced using standard routines in ATNF's MIRIAD package \citep{1995ASPC...77..433S}. Primary flux calibration for all observations was done using 1934$-$638. Phase calibration was carried out with 1817$-$254 during the 2011 September 05, 20, and 21 observations. 1908$-$201 was used for all other observations, except on 2012 April 08 where 1829$-$207 was used. Due to a combination of antenna shadowing in the compact H75 configuration and a missing antenna, only one baseline was usable for the 2011 September 20 and 21 observations. Therefore, for these two epochs a point source was fit to the calibrated data in the uv-plane. For all other observations, calibrated data were then exported to \textsc{Difmap} for imaging and flux densities were obtained by fitting a point source model to the target in the image plane (Table~\ref{tab:ATCA_flux}). Due to the compact array configuration on 2011 September 20 and 21, our ATCA observations were unable to resolve \source{} from a nearby unrelated object. The flux density of the nearby source was subtracted from our measured flux density to estimate the flux density of \source{} in our ATCA observations (see Section~\ref{sec:VLA} and Figure~\ref{fig:source} for further discussion).

The radio counterpart of \source{} was detected for all 2011 observations at a position coincident with those of the X-ray and optical counterparts. During our 2012 observations, the source was detected only on April 08 (during the minor re-flare). When detected, the source was found to be unresolved within the ATCA beam.

\begin{table}
\caption{ATCA flux densities of \source{}. Due to the presence of a nearby unrelated source (within the ATCA beam, Figure~\ref{fig:source} inset), for the observations in which the array was in the compact H75 configuration the interpolated flux density of the unrelated source (determined from our VLA observations; Figure~\ref{fig:source}) was subtracted to estimate the flux density of \source{} (see Section~\ref{sec:VLA}). At other epochs, the more extended array configuration meant the two sources were resolved from each other. Upper limits are 3$\sigma$. }
\centering
\label{tab:ATCA_flux}
\begin{tabular}{ccccccc}
\hline
Date & MJD & Array & Freq. & Flux \\
 (UT) & & config. & (GHz) & density \\
& & & & (mJy) \\
\hline
2011 Sep 05 & 55809.45 & 6B & 5.5 & 28$\pm$2\\
 & &  & 9.0 & 36$\pm$4\\
2011 Sep 20 & 55824.22 & H75 & 5.5 & 45$\pm$4$^{\footnotesize{\rm a}}$\\
 & &  & 9.0 & 50$\pm$3$^{\footnotesize{\rm a}}$\\
2011 Sep 21 & 55825.21 & H75 & 5.5 & 39$\pm$4$^{\footnotesize{\rm a}}$\\
 & &  & 9.0 & 44$\pm$3$^{\footnotesize{\rm a}}$\\
2011 Dec 01 & 55897.27 & 1.5D & 5.5  & 0.4$\pm$0.1\\
 & &  & 9.0 & 0.24$\pm$0.05\\
2012 Jan 13 & 55939.92 & 6A & 5.5 & $\le$0.09 \\
 & &  & 9.0 & 0.27$\pm$0.07\\
2012 Feb 03 & 55960.05 & 6A & 5.5 & $\le$0.11 \\
 & &  & 9.0 & $\le$0.21\\
2012 Apr 08 & 56024.94 & H168 & 5.5 & 0.64$\pm$0.05\\
 & &  & 9.0 & 0.68$\pm$0.04\\
2012 Jun 24 & 56102.73 & 6D & 5.5 & $\le$0.23\\
 & &  & 9.0 & $\le$0.28\\
2012 Aug 24 & 56163.30 & 6A & 5.5 & $\le$0.07\\
 & &  & 9.0 & $\le$0.09\\
\hline
\multicolumn{5}{l}{Preliminary results for the 2012 April 08 epoch were reported}\\
\multicolumn{5}{l}{by \citet{2012ATel.4038....1C}.}\\
\multicolumn{5}{l}{$^{\footnotesize{\rm a}}$The flux density of the nearby source was subtracted from our}\\
\multicolumn{5}{l}{measured flux density (see Section~\ref{sec:VLA}).}\\
\end{tabular}
\end{table}

\subsection{X-ray observations}
\subsubsection{RXTE}
\begin{figure}
\centering
\includegraphics[width=\columnwidth]{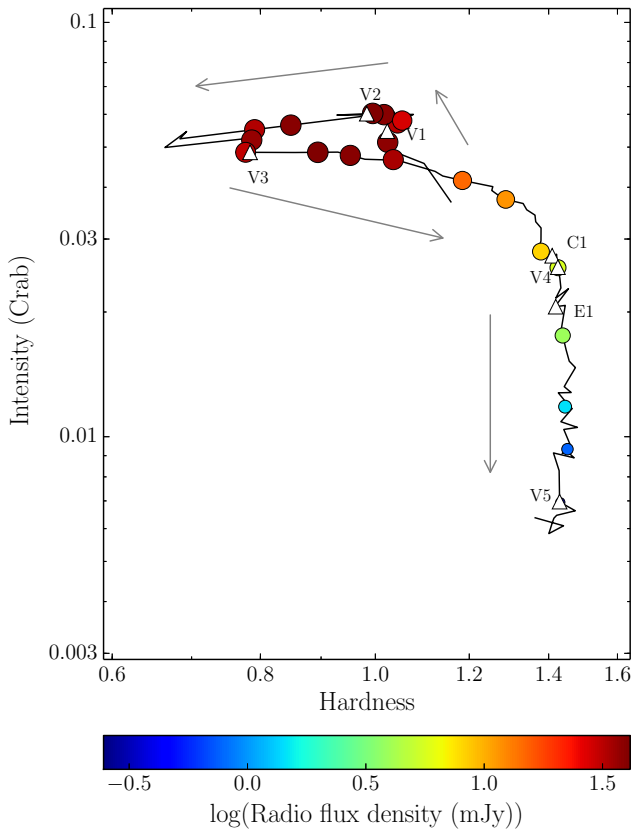}
\caption{Hardness-intensity diagram of MAXI J1836-194. The {\it RXTE} X-ray hardness (6--16 keV/2--6 keV) and intensity (2--20 keV) are normalised to the Crab. The 5\,GHz VLA and 5.5\,GHz ATCA radio observations are shown as filled circles, with the marker size and colour representing the radio flux density. The white triangles represent the timing of the VLBI observations (including the two CVN and EVN observations from \citealt{2012MNRAS.426L..66Y}, see Section~\ref{sec:pm}). V1 through V5 refer to the VLBA observational epochs, C1 is the CVN observation and E1 is the EVN observation. Our high-cadence radio observations show the full evolution of the compact jet during the failed outburst.}
\label{fig:hid}
\end{figure}

The outburst of \source{} was monitored with the Proportional Counter Array (PCA) on board the {\it Rossi X-ray Timing Explorer} ({\it RXTE}; see \citealt{1996SPIE.2808...59J,2006ApJS..163..401J}). A total of 74 pointed observations sample this outburst, for a period of about 92 days. The {\it RXTE} data were reduced using the Heasoft software package v6.8, following the standard steps described in the {\it RXTE} cookbook. To estimate the source intensity and a spectral colour, we followed \citet{2008ApJ...685..436A}: we used the 16 second time-resolution Standard 2 mode of all active Proportional Counter Units (PCUs, all layers) in each observation. The hard colour was defined as the count rate ratio (6--16 keV)/(2--6 keV) and the intensity was defined as total count rate in the 2.0--20.0 keV band (Figure~\ref{fig:hid}). We subtracted the background and performed deadtime corrections on a per-PCU basis. To take into account response differences between PCUs, we normalised the intensity and hard colour by the Crab before estimating the average intensity and hard colour on a per-observation basis. Due to the low column density along the line-of-sight ($n_{\rm H} = 3.0 \times 10^{21}$ cm$^{-2}$; \citealt{2014MNRAS.439.1390R}) and the spectrum being dominated by a power-law component during the hard state decay, the 3--9 keV band during the decay phase of the outburst was unaffected by absorption.  

\begin{figure*}
\centering
\includegraphics[width=\textwidth]{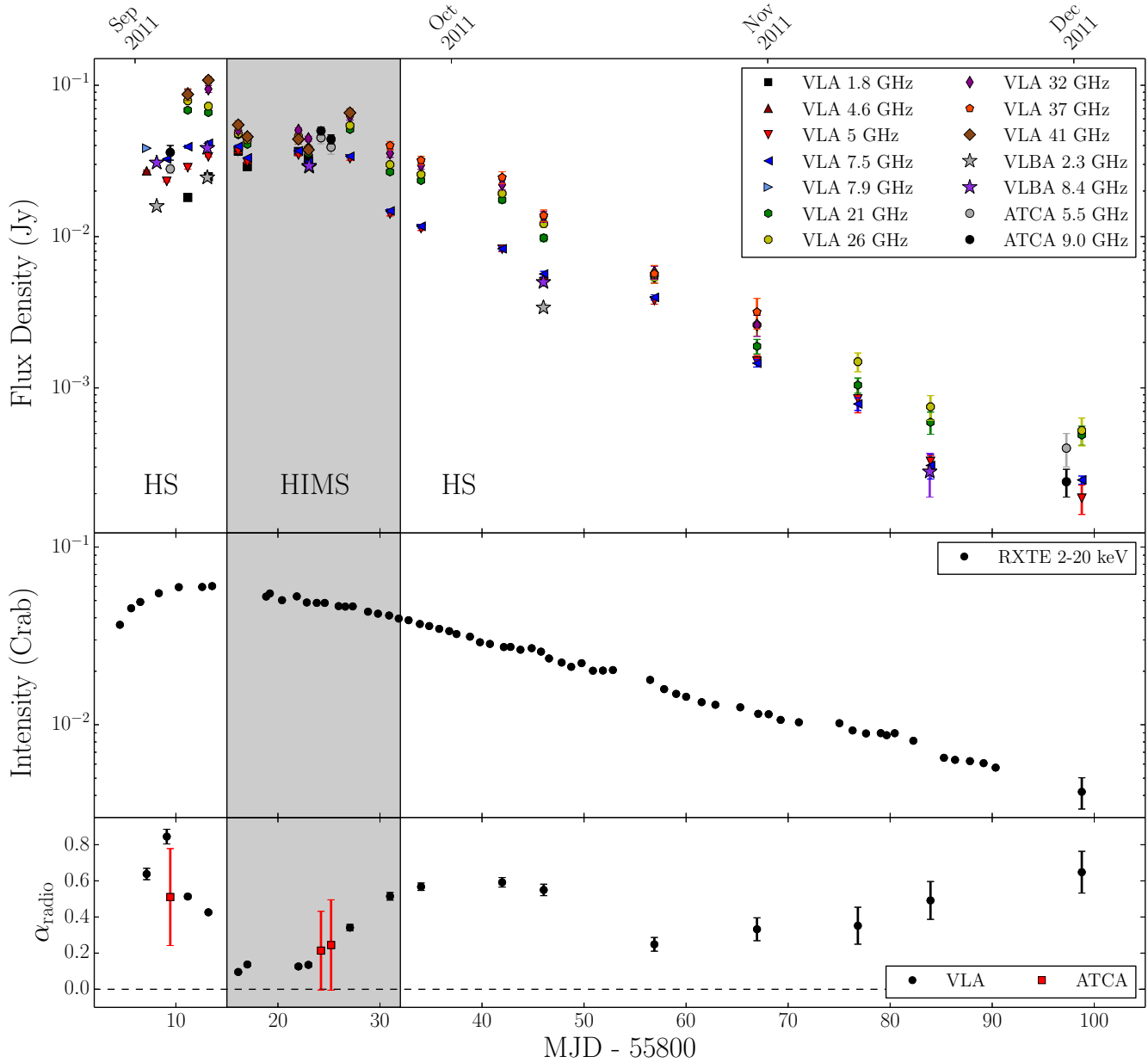}
\caption{Radio and X-ray lightcurves of \source{} during the VLA, VLBA, ATCA and RXTE monitoring. The grey shaded region represents the HIMS (also denoted by HIMS) and HS is the hard state, as defined by \citet{ferrigno2012}. The top panel shows the flux densities of the VLA, VLBA and ATCA monitoring. The second panel gives the RXTE 2--20 keV lightcurve (normalised to the Crab) and the third panel shows the evolution of the radio spectral index, $\alpha$, where $S_{\nu} \propto \nu^{\alpha}$ (calculated from the ATCA and VLA observations, see Section~\ref{sec:spectral_indices}). The spectral index flattens during the HIMS, and then becomes more inverted during the decay.}
\label{fig:lightcurves}
\end{figure*}

Due to Sun constraints, no simultaneous RXTE data were available to coincide with the 2011 December 3 VLA data (the final RXTE epoch was 2011 November 25). We estimate the X-ray intensity on December 3 by fitting the observed exponential decay of the source following its transition back to the hard state (see Figure~\ref{fig:lightcurves}), and extrapolating to the required date (error bars were conservatively estimated from the uncertainty of the exponential fit).

\subsection{{\it Swift} XRT}

The \textit{Swift} X-ray telescope (XRT) monitored \source{} during its period of renewed activity in 2012.  From the HEASARC public archives, we retrieved XRT observations taken on 2012 April 06 and 09, June 15 and 29 (which were closest in time to our April 08 and June 24 ATCA observations). The 2012 April observations (characterised by a higher count rate) were taken in windowed-timing mode, while those from 2012 June (when the source was fainter) were in photon-counting mode. We extracted the spectral files and their associated background, response and ancillary response files with the online XRT data product generator \citep{2009MNRAS.397.1177E}. The X-ray spectra were then fit with the X-Ray Spectral Fitting Package ({\small XSPEC}) version 12.7 \citep{1996ASPC..101...17A}. We modelled the spectra in the 0.4--10 keV band with a simple disk-blackbody plus power-law, absorbed by a (fixed) line-of-sight Galactic $n_{\rm H} = 3.0 \times 10^{21}$ cm$^{-2}$ \citep{2014MNRAS.439.1390R} and free intrinsic absorption. More complex fits with physical Comptonisation
models or irradiated disk models did not improve the fit due to the low signal-to-noise ratio of our data. Moreover, our primary interest was in the flux from the power-law component in the 3--9 keV band, rather than the details of the soft component. Therefore, a disk-blackbody plus power-law model was adequate. On 2012 April 06 and 09, a faint disk-blackbody component was significantly detected with a temperature $kT_{\rm in}=0.16^{+0.04}_{-0.02}$ keV (from the combined spectra), but the 3--9 keV flux was entirely dominated by the power-law component ($\Gamma = 1.65^{+0.10}_{-0.08}$) and was unaffected by absorption (total $n_{\rm H} \la 4 \times 10^{21}$ cm$^{-2}$). On 2012 June 15, the spectrum was best fit by a simple power law ($\Gamma = 1.98^{+0.35}_{-0.32}$). On 2012 June 29, the source was no longer significantly detected. We therefore estimated a 90\% confidence limit to its net count rate from the number of counts detected in the source and background regions (using Bayesian statistics; \citealt{1991ApJ...374..344K}). We converted this count rate upper limit to a flux upper limit assuming the same spectral model that we fit to the 2012 June 15 spectrum.


\begin{figure}
\centering
\includegraphics[width=\columnwidth]{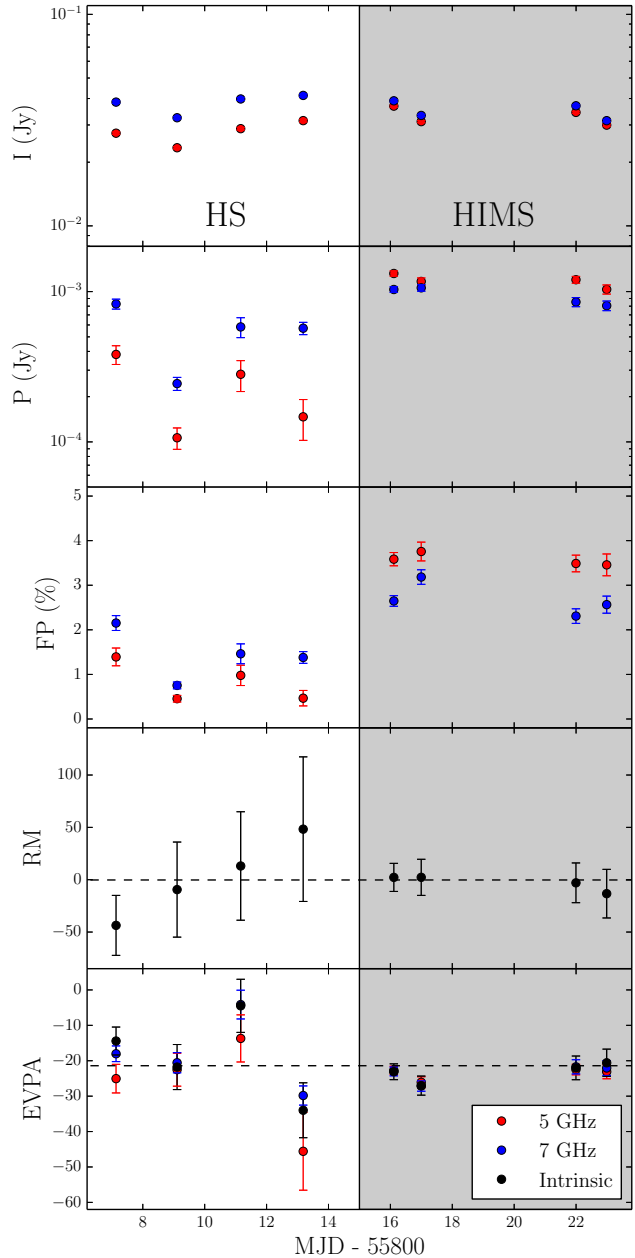}
\caption{Polarisation parameters of \source{} during the early part of the outburst, before the VLA shifted to a more compact configuration and the target could not be resolved from the nearby confusing source. The unshaded region represents the hard state (HS) and the shaded represents the HIMS. Top panel: Stokes $I$ at 5 and 7.5\,GHz (red and blue points respectively). Second panel: linear polarisation ({\it P}) at 5 and 7.5\,GHz (red and blue points respectively), where {\it P}$= \sqrt{Q^2 + U^2}$. Third panel: fractional polarisation ({\it FP}) at 5 and 7.5\,GHz (red and blue points respectively), where {\it FP}$= 100 \times P / I$. Fourth panel: the rotation measure ({\it RM}), the black dashed line is a weighted mean of the {\it RM} values. Fifth panel: the intrinsic electric vector position angle ({\it EVPA}; black circles), derived from fitting the observed 5 and 7.5\,GHz polarisation angles on a per-subband basis against $\lambda^2$. The red and blue points represent the 5 and 7.5\,GHz polarisation angles (per-baseband), respectively. We see low levels of fractional polarisation from \source{} during the early hard state and HIMS, a steady {\it RM} (of $|{\it RM}|\le 32$ rad m$^{-2}$) and weighted mean {\it EVPA} of $-21\pm4^{\circ}$ (fifth panel, black dashed line).}
\label{fig:polarisation}
\end{figure}

\section{Results}
\label{sec:results}

\subsection{Source evolution}
\label{sec:spectral_indices}

With our VLA, VLBA, and ATCA monitoring, we observed the brightening and decay of the compact jet of \source{} during its 2011 outburst (Figure~\ref{fig:lightcurves}, top panel). The VLA, VLBA and ATCA radio observations (Tables~\ref{tab:VLA_flux},~\ref{tab:xband_fits} and~\ref{tab:ATCA_flux}, respectively) show that the flux density of the system increased during the bright, early hard state and HIMS, before the outburst failed. The VLA radio spectrum (which was initially inverted in the hard state) flattened as the system softened and transitioned to the HIMS (Figure~\ref{fig:lightcurves}, bottom panel). Never entering the soft state, the source transitioned back to the hard state on September 28 and the radio emission faded by $\sim$2 orders of magnitude and the spectrum became more inverted during the decay towards quiescence. 

The X-ray evolution of a black hole X-ray binary is often visualised with a hardness intensity diagram (HID), which shows the variation of X-ray intensity with X-ray spectral shape (where harder spectra are dominated by the X-ray power-law component and the softer emission by the blackbody disk emission; e.g. \citealt{2004MNRAS.355.1105F}). Figure~\ref{fig:hid} shows how the flux density of the VLA and ATCA observations varied with X-ray intensity and spectral shape, and indicates the timing of the VLA, ATCA and VLBA radio observations.

\begin{figure*}
\centering
\includegraphics[width=\textwidth]{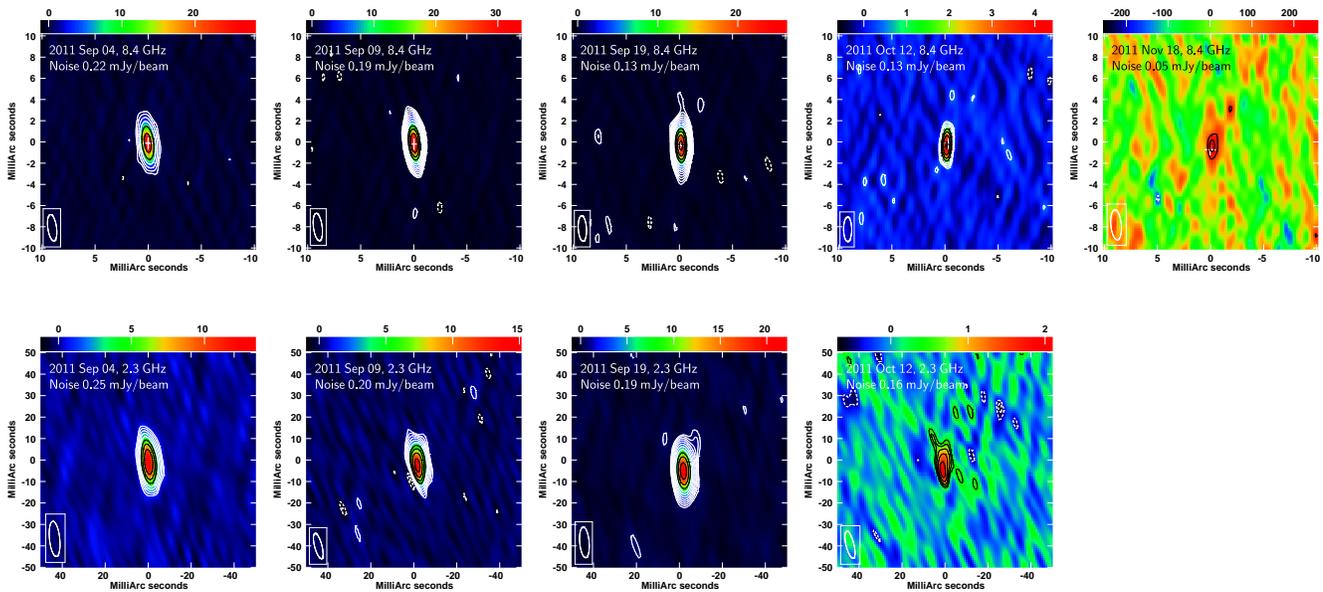}
\caption{VLBA images of MAXI J1836-194. Top panels are at 8.4\,GHz, and bottom panels are at 2.3\,GHz. Contour levels are $\pm2^n\sqrt{2}$~times the rms noise level indicated in the top left corner of each image, where $n=-2,2,3,4,...$. The colour scale shows the image flux density in units of mJy\perbeam, except for the final 8.4-GHz image (2011 November 18), for which the units are $\mu$Jy\perbeam.  The white crosses in the upper panels mark the predicted core position from the proper motion fit performed in Section~\ref{sec:pm}. The source is unresolved except in the 2.3-GHz image of 2011 September 19, where a $\sim$15\,mas extension is seen along the position angle $-29\pm3^\circ$.}
\label{fig:VLBA_images}
\end{figure*}

\subsection{Polarisation of the radio jet}
\label{sec:results_polarisation}

Linear polarisation was significantly detected ($>3\sigma$) for the first eight VLA epochs at both 5 and 7.5\,GHz (see Figure~\ref{fig:polarisation}, second panel). The linear polarisation ({\it P}), fractional polarisation ({\it FP}) and polarisation position angle ({\it PA}) were derived from the measured flux densities of the Stokes $Q$ and $U$ images, where {\it P}$= \sqrt{Q^2 + U^2}$, {\it FP}$= 100 \times P / I$ and {\it PA}$= 0.5 \arctan{(U/Q)}$. Following the September 18 epoch the target faded and the VLA shifted to a more compact configuration and, due to the presence of the polarised confusing source, we were not able to place any constraints on the source polarisation.  

Faraday rotation is the rotation of the plane of polarisation as a function of wavelength ($\lambda$), due to birefringence of the interstellar and local medium. The rotation measure, {\it RM}, is given by $RM \propto \int_0^d n_{\rm e} B_{\rm ||} {\rm d}l$ (where $n_{\rm e}$ is the thermal electron density along the line of sight, $B_{\rm ||}$ is the magnetic field along the line of sight, and $d$ is the source distance). The intrinsic electric vector position angle ({\it EVPA}) of the source is related to the wavelength and polarisation position angle by {\it EVPA}$=${\it PA}$-${\it RM}$\lambda^2$. Therefore, the {\it RM} and {\it EVPA} can be determined from the 5 and 7.5\,GHz observations by linearly fitting the {\it PA} for all 16 subbands (ensuring we do not suffer from an $n\pi$ ambiguity). Our results show that the {\it RM} and {\it EVPA} remained constant within errors throughout the outburst and we find that $|{\it RM}|\le 32$ rad m$^{-2}$ and {\it EVPA}$=-21\pm4^{\circ}$ (Figure~\ref{fig:polarisation}, fourth and fifth panel, respectively).

\subsection{VLBA radio morphology}
\label{sec:vlbi_axis}

Our VLA, ATCA and VLBA radio observations show similar flux densities of \source{} during its outburst. Also, contemporaneous VLBI observations of \source{} with the CVN (on 2011 October 10) and the EVN (on 2011 October 17) reported by \citet{2012MNRAS.426L..66Y}, show similar flux densities to our contempraneous radio observations, suggesting that the high resolution observations do not resolve out any significant extended emission.

\begin{table*}
\caption{VLBA positions of MAXI J1836$-$194 at 8.4 GHz at each epoch. We see the fitted position change by 0.76$\pm$0.09\,mas in right ascension (R.A.) and 0.9$\pm$0.2\,mas in declination (Dec.). }
\centering
\label{tab:VLBA_positions}
\begin{tabular}{ccccc}
\hline
Date & MJD & Freq. & R.A. & Dec.\\
2011 &  & (GHz) & &\\
(UT) &  &       &  & \\
\hline
Sep 04 & $55808.12\pm0.06$ & 8.42 &  18:35:43.4445732(8) & -19:19:10.48441(3)\\
Sep 09 & $55813.08\pm0.08$ & 8.42 &  18:35:43.4445671(5) & -19:19:10.48438(2)\\
Sep 19 & $55823.06\pm0.08$ & 8.42 &  18:35:43.4445587(4) & -19:19:10.48470(2)\\
Oct 12 & $55846.02\pm0.10$ & 8.42 & 18:35:43.444564(1)\phantom{0}  & -19:19:10.48520(4)\\
Nov 18 & $55883.88\pm0.10$ & 8.42 & 18:35:43.444519(6)\phantom{0}  & -19:19:10.4853(2)\phantom{0}\\
\hline
\end{tabular}
\end{table*}

The compact jet was not directly resolved in the individual images, except on September 19 at 2.3~GHz. Model fitting in the {\it uv}-plane shows the source to be slightly extended at all frequencies, and in all but the last epoch (Table~\ref{tab:xband_fits}), as confirmed during the deconvolution process while imaging. As a check on the quality of the fits, the fitted Gaussian was subtracted from the {\it uv}-data, and the data were re-imaged. The resulting residual images were noise-like, with the exception of the 2.3~GHz September 19 observation, for which some structure remained at the level of 1.0\,mJy\,beam$^{-1}$, suggesting that a simple elliptical Gaussian was not a good representation and that the source structure was more complex (Figure~\ref{fig:VLBA_images}). However, since it can account for $>$96\% of the emission at that epoch, we take the fitted parameters as a good approximation to the source structure in this epoch, together with a faint residual jet component separated by $\sim$15\,mas from the core, along the position angle $-29\pm3^\circ$. Excluding the September 04 2.27\,GHz observation, which was taken early in the outburst with relatively low resolution, model fitting in the {\it uv}-plane gave a consistent position angle, with a weighted mean of $-21\pm2^\circ$. Hereafter, we take the weighted mean as indicative of the observed jet axis. The fitted major axis of the source (which was smaller at higher-frequency, corresponding to smaller scales within the jet; see also \citealt{2012MNRAS.419.3194R}) did not change significantly during the outburst (Table~\ref{tab:xband_fits}).

\subsection{Astrometry}
\label{sec:pm}

To constrain the proper motion of the system, we measured the source position in each of the 8.4\,GHz images prior to performing any self-calibration (Table~\ref{tab:VLBA_positions}). In all cases, we fitted the source in the image plane with an elliptical Gaussian, for consistency with the known source structure. We estimated systematics from the scatter on the two check source positions over the five epochs, weighted by their relative distance to the phase calibrator. Over the two months of our observing campaign, the fitted source position changed by 0.76$\pm$0.09\,mas in right ascension (R.A.) and 0.9$\pm$0.2\,mas in declination (Dec.). To increase the size of our data set, we included the 8.3-GHz position derived by \citet{2012MNRAS.426L..66Y} from their Chinese VLBI Network (CVN) observations. Although the CVN observations used the same phase reference source, J1832-2039, they used a less recent global astrometric solution to determine the assumed position of that calibrator, so we corrected for the shift in calibrator position before adding the CVN point to our astrometric sample.

We also attempt to correct for the motion of the optical depth $\tau=1$ surface along the jet axis, by calculating the expected core shift as a fraction of the component of proper motion along the VLBI jet axis. \citet{2006ApJ...636..316H} present a model for the expected core shift in a compact jet. Assuming equipartition between the particles and the magnetic field and following their assumptions, we calculate the expected core shift at each epoch. We find a cumulative shift that is $\la$15\% of the total positional shift along the jet axis, which we include (on a point by point basis) in the uncertainties of our proper motion fit.

\begin{figure}
\centering
\includegraphics[width=\columnwidth]{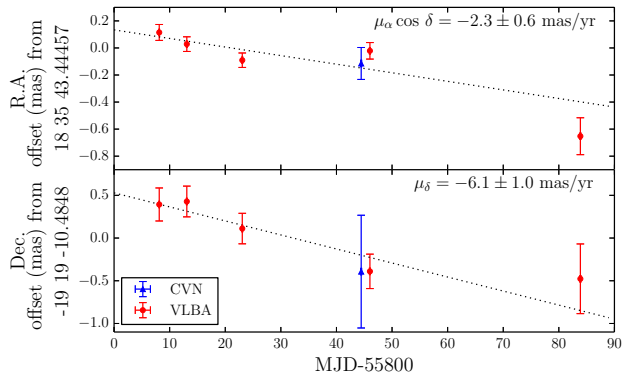}
\caption{The measured positions of \source{} as a function of time (positions given in Table~\ref{tab:xband_fits}). Here, the motion in R.A. and Dec. are shown in the upper and lower panels, respectively. VLBA measurements are shown in red, and the 8.3-GHz CVN measurement of \citet{2012MNRAS.426L..66Y} is shown in blue. We include the systematic uncertainty determined from the scatter in the check source positions, as well as the uncertainty due to the core shift of the jet in our error bars (applied on a point by point basis). These high-resolution observations show the proper motion of the system.}
\label{fig:pm}
\end{figure}

Using jackknife resampling with a linear function to fit the motion in both co-ordinates, we derived proper motions in R.A. and Dec. of
\begin{align}
\mu_{\alpha}\cos\delta &= -2.3\pm0.6\quad {\rm mas}\quad{\rm year}^{-1}\\
\mu_{\delta} &= -6.1\pm1.0\quad {\rm mas}\quad{\rm year}^{-1},
\end{align}
with the reference position
\begin{align}
{\rm R.A.} &= 18^{\rm h}35^{\rm m}43\fs444575 \pm 0.000003\\
{\rm Dec.} &= -19^{\circ}19^{\rm m}10\farcs4843 \pm 0.0002
\end{align}
at the reference date of MJD 55800.0 (Table~\ref{tab:xband_fits} and Figure~\ref{fig:pm}).

\section{Discussion}

\subsection{The evolving compact jet}
\label{sec:evolution}

During the initial radio observations of \source{}, the system was in the hard X-ray spectral state. In this state, the radio spectrum was very inverted (Figure~\ref{fig:lightcurves}, bottom panel) and linear polarisation was observed at levels of a few percent (Figure~\ref{fig:polarisation}), significantly lower than the maximum expected theoretical levels for a perfectly ordered magnetic field. Synchrotron theory predicts linear polarisation at levels of up to $\sim$10\% for optically-thick synchrotron emission and up to $\sim$72\% for optically-thin synchrotron emission in the presence of an ordered magnetic field \citep{1967ApJ...150..647P,1968ARA&A...6..321S,1979rpa..book.....R,2011hea..book.....L}. However, these levels have rarely been observed in LMXBs \citep{2003Ap&SS.288...79F}. For example, like \source{}, linear polarisation at levels of a few percent was observed in the hard states of V404~Cyg \citep{1992ApJ...400..304H} and GX~339$-$4 \citep{2000A&A...359..251C}. The low polarisation implies either that the magnetic field was intrinsically disordered on large scales \citep[e.g.][]{2007MNRAS.378.1111B}, possibly due to multiple polarised components (which are polarised differently; \citealt{2004MNRAS.354.1239S}), a rotation of the polarisation angle within the jet \citep{1979ApJ...232...34B} or Faraday rotation within the jet (see \citealt{2013MNRAS.432..931B} for further discussion). 

As the system transitioned to the HIMS, the radio emission brightened, the radio spectrum flattened, and the linear polarisation remained at a few percent (although it did increase marginally). During this time, the polarisation spectrum steepened slightly (Figure~\ref{fig:polarisation}, second panel), which the evolution of {\it FP} shows is not just due to the change in the spectral index of Stokes~{\it I}. While our observations do not allow for a clear interpretation of this result, we speculate that this steepening may have been due to the magnetic field becoming intrinsically more ordered further along the jet (away from the compact object), or less ordered closer to the compact object. The outburst then failed in the HIMS and the radio spectrum became more inverted as the source transitioned back to the hard state. 

\begin{figure}
\centering
\includegraphics[width=\columnwidth]{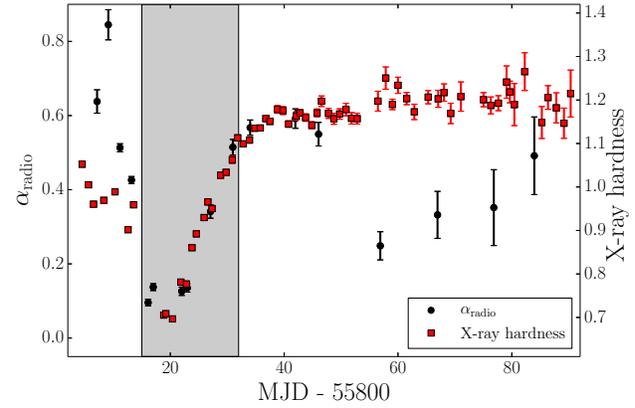}
\caption{The VLA spectral index ($\alpha$) and {\it RXTE}/PCA hardness ratio (6-16\,keV/2-6\,keV) of \source{} during its 2011 outburst. $\alpha$ (left hand axis) is represented by the black points and X-ray hardness (right axis) is shown as the red squares. The grey shaded region represents the HIMS. During the HIMS and transition back to the HS $\alpha$ appeared to track the X-ray hardness. Following the transition back to the hard state the X-ray hardness remained steady while the jet continued to evolve.}
\label{fig:alpha_Gamma}
\end{figure}

During the initial hard state observations of \source{}, the radio spectrum was very inverted, with a spectral index of $\alpha \sim 0.6-0.8$ (determined from the Stokes~{\it I}). While X-ray binaries in the hard state typically exhibit flat or slightly inverted radio spectra, where $\alpha \sim 0-0.6$ \citep[e.g. ][]{2001MNRAS.322...31F,2009ApJ...698.1398M,2013MNRAS.429..815R,2013MNRAS.432..931B,2013MNRAS.436.2625V,2014MNRAS.437.3265C}, inversions of $\alpha \sim 0.7$ were observed during the decay phase of the 2010-2011 outburst of GX~339$-$4 \citep{2013MNRAS.431L.107C}. The radio spectrum of \source{} then flattened (to $\alpha \sim 0.1-0.2$) as the system transitioned into the HIMS. Our results show that during the transition phases of the outburst, the radio spectrum appeared to track the X-ray hardness (Figure~\ref{fig:alpha_Gamma}). However, during the hard state decay the spectrum of the radio jet continued to evolve while the X-ray hardness remained steady. During this time, the jet spectral break was observed to change by an order of magnitude indicating that the jet properties are still changing significantly (\citealt{2013ApJ...768L..35R}; \citealt{2013ApJ...779...95K}; \citealt{2014MNRAS.439.1390R}). 

A flattening of the radio spectrum as the source softens has been observed in some (but not all) other systems. The shape of the spectrum is thought to be dependent on the geometry of the compact jet and the flow of material along the jet (e.g. the bulk flow velocity along the jet and shocks and turbulence within the compact jet; \citealt{1979ApJ...232...34B,1995A&A...293..665F,2006MNRAS.367.1083K,2009ApJ...699.1919P, 2010MNRAS.401..394J,2013MNRAS.429L..20M,2014MNRAS.443..299M}). At the beginning of its 2003 outburst, the radio spectrum of H~1743$-$322 showed a similar evolution to \source{}, where the inverted spectrum of the compact jet flattened as the system softened during its transition to the intermediate state \citep{2009ApJ...698.1398M}. \source{} reached its softest point in the HIMS, following which the jet spectrum became more inverted as the system transitioned back to the hard state. GX~339-4 \citep{2013MNRAS.431L.107C} showed a similar evolution where the radio spectrum became more inverted following the transition back to the hard state (after the compact jet was re-established). However, a radio spectrum that steepens as the source softens is not observed in all LMXBs \citep[e.g.][]{2013MNRAS.436.2625V,2014MNRAS.437.3265C}. From this small sample, the difference in the evolution of the jet spectrum between systems does not appear to be related to any physical parameter of the system such as inclination angle and black hole mass. While it appears that the evolution of the radio spectrum must be driven by some property of the accretion flow (that may also be driving X-ray hardness), the surrounding environment, or the magnetic field (that may affect the jet flow and shape), our observations do not allow us to identify what processes may be driving the evolution.

Interestingly, while the flux density of the target changed, the fitted size of the compact jet at 8\,GHz remained relatively constant (Table~\ref{tab:xband_fits}). According to \citet{2006ApJ...636..316H}, the distance of the radio core along the jet axis, $z_{\rm 0}$, is expected to scale with the observed flux density, $S_{\nu}$, where $z_{\rm 0} \propto S_{\nu}^{8/17}$. The size of the compact jet, $d_{\rm core}$, is expected to scale linearly with the distance of the radio core from the origin \citep{2013MNRAS.432.1319P}. Therefore, the fitted size of the compact jet should relate to the flux density of the system, where $d_{\rm core} \propto S_{\nu}^{8/17}$. In our VLBA observations we therefore expect to see the jet change size by a factor of $\sim$2.5 at 8\,GHz. However, our results show that the jet size remained steady during the outburst (even as the jet faded significantly), which could be due to physical properties of the jet changing as it faded (such as changes to the opening angle, jet confinement region etc.). 

The VLA observations show that the {\it RM} remained approximately constant throughout the observations (Figure~\ref{fig:polarisation}), where $|${\it RM}$|~ \le$32~rad\,m$^{-2}$ (to 1$\sigma$). This result is consistent with values found by \citet{2009ApJ...702.1230T} for radio sources along a similar line of sight (where $-100~\la$~{\it RM}~$\la$~100~rad\,m$^{-2}$). Faraday rotation of the linearly polarised synchrotron emission is related to the thermal electron density and magnetic field along the line of sight. Therefore, the stable {\it RM} suggests that the rotation is not dominated by local material (to the source). The low value of the RM suggests that Faraday rotation is most likely not responsible for the low levels of observed linear polarisation.

\subsection{Jet axis alignment}

We find an agreement between the calculated EVPA ($-21~\pm~4^{\circ}$; Section~\ref{sec:results_polarisation}) and the observed VLBI jet axis ($-21~\pm2^{\circ}$; Section~\ref{sec:vlbi_axis}). The total jet spectrum is composed of contributions from both the optically-thick and optically-thin regions of each of the individual synchrotron components \citep{1979ApJ...232...34B,1988ApJ...328..600H}. Fractional polarisation is considerably higher in the optically-thin region of each synchrotron component than the optically-thick region; hence, it is expected that the observed polarisation is dominated by contributions from the optically-thin region of each synchrotron component \citep{2014MNRAS.442.3243Z} and therefore that the EVPA will be aligned perpendicular to the magnetic field. This implies that the magnetic field is perpendicular to the jet axis in \source{}. A perpendicular magnetic field may arise from shock compression of the magnetic field \citep{1980MNRAS.193..439L}, knots or hotspots in the jet \citep{1987ApJ...316..611D}, or a helical magnetic field structure dominating the observed emission \citep{2008Natur.452..966M}, a consequence of the rotation of the accretion disk from which the jets may be launched \citep{2002Sci...295.1688K,2007MNRAS.380...51K,2008Natur.452..966M,2008ApJ...685..333G}. A similar alignment between jet axis and EVPA has also been observed in the black hole X-ray binaries GX 339$-$4 \citep{2000A&A...359..251C}, GRO J1655-40 \citep{2000ApJ...540..521H} and Cygnus X$-$1 \citep[][although this measurement is model-dependent]{2014MNRAS.438.2083R}. Polarisation studies of BL Lac sources have shown a bimodal distribution of the EVPA relative to the jet axis, where the majority of sources show an alignment between the EVPA and the jet axis (e.g. \citealt{2014ApJ...787..151C} and references therein). If an alignment between the EVPA and jet axis is common to black hole X-ray binaries, high resolution images would not be required to determine the orientation of the jets.

\subsection{Proper motion}

\begin{figure}
\centering
\includegraphics[width=\columnwidth]{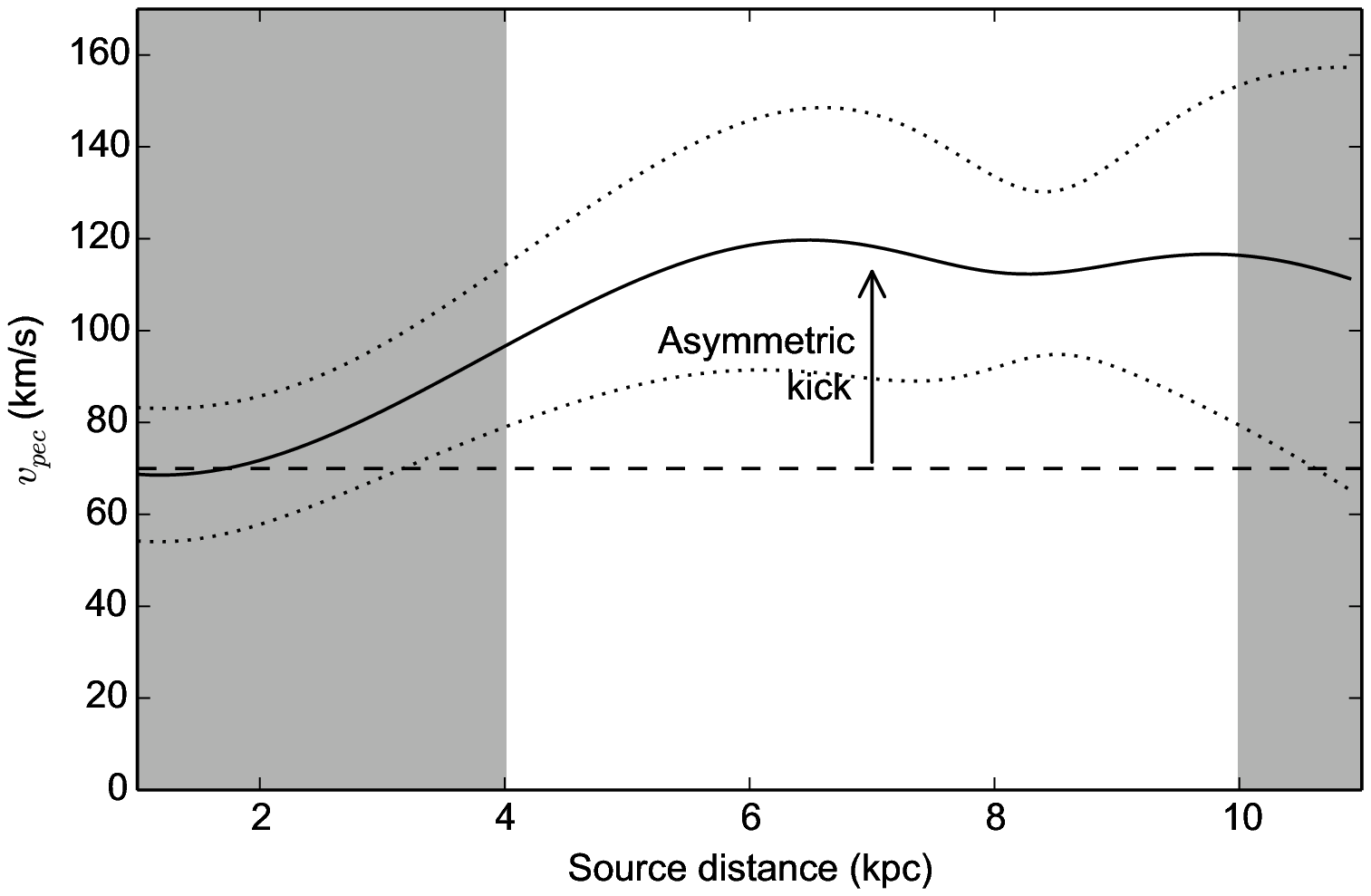}
\caption{Peculiar velocity ($v_{\rm pec}$) of \source{} as a function of source distance (in kpc). The black solid line is $v_{\rm pec}$, the dotted lines are the error bars from uncertainties in proper motion and systemic velocity, and the unshaded regions denote the source distance limit of between 4 and 10 kpc \citep{2014MNRAS.439.1390R}. The area above the horizontal dashed line represents the requirement of an asymmetric supernova kick. Our results suggest that an asymmetric supernova kick is required to explain the peculiar velocity of the system.}
\label{fig:vpec}
\end{figure}

From this single outburst we calculate the proper motion of the system by linearly fitting its VLBI positions in both co-ordinates over time (see Section~\ref{sec:pm} and Figure~\ref{fig:pm}), which can then be used to constrain the formation mechanism of the black hole. Black holes in LMXB systems are thought to form in two ways \citep{2001ApJ...554..548F}: either a massive star collapsing directly into a black hole \citep[e.g.][]{2003Sci...300.1119M,2007ApJ...668..430D}, or delayed formation in a supernova, as fallback on to the neutron star of material ejected during the explosion creates the black hole \citep[e.g.][]{2001Natur.413..139M}. Following a supernova, the centre of mass of the ejected material continues moving with the velocity of the progenitor immediately prior to the explosion. The centre of mass of the binary system then recoils in the opposite direction and is constrained to lie in the orbital plane (Blaauw kick; \citealt{1961BAN....15..265B}). In the presence of asymmetries in the supernova explosion, a further asymmetric kick, which need not be in the orbital plane, may be imparted to the binary \citep[e.g.][]{1995MNRAS.274..461B,1998A&A...332..173P,2001ApJ...549.1111L,2005ApJ...618..845G,2005ApJ...625..324W,2009ApJ...697.1057F}. For further discussion on formation see \citet{2014PASA...31...16M} and references therein. 

With our calculated proper motion, source distance (4--10 kpc; \citealt{2014MNRAS.439.1390R}), and systemic radial velocity (61$\pm$15 km s$^{-1}$; \citealt{2014MNRAS.439.1381R}), we can determine the peculiar velocity of this system. We calculate the heliocentric Galactic space velocity components using the formalism of \citet{1987AJ.....93..864J}, assuming the standard solar motion measured by \citet{2010MNRAS.403.1829S}. We assume circular rotation about the Galactic Centre in the plane of the disk, a flat rotation curve with a tangential velocity of 240\,km\,s$^{-1}$, and a Galactocentric distance for the Sun of 8.34\,kpc \citep{2014ApJ...783..130R}.  Defining the peculiar velocity as the difference between the measured three-dimensional space velocity and the expected motion of the system due to Galactic rotation \citep[see, e.g.][]{2007ApJ...668..430D}, we find that the peculiar velocity is $>70$\,km\,s$^{-1}$ (Figure~\ref{fig:vpec}). Since stellar velocity dispersion in the disk gives a typical peculiar velocity of $<45$\,km\,s$^{-1}$ even for the oldest M-type stars \citep{2000A&A...354..522M}, and typical Blaauw kicks in LMXBs (while system parameter dependent) have a maximum recoil velocity of $\sim70$\,km\,s$^{-1}$ \citep[e.g.][]{1999A&A...352L..87N}, an asymmetric natal kick was most likely required in \source{}. 

This is just the sixth system in which constraints have been placed on the formation mechanism of the black hole \citep[see][and references therein]{2014PASA...31...16M}. While measuring the distribution of black hole kick velocities can constrain the formation mechanism of black holes, these observations must be carried out in a quiescent or hard state, which are faint (high resolution observations of LMXBs are generally taken during full outburst). Therefore, with so few systems having accurate constraints, there is relatively little observational data on black hole kicks and observations such as these are required to make further progress.

\subsection{Radio/X-ray correlation}

\begin{figure}
\centering
\includegraphics[width=\columnwidth]{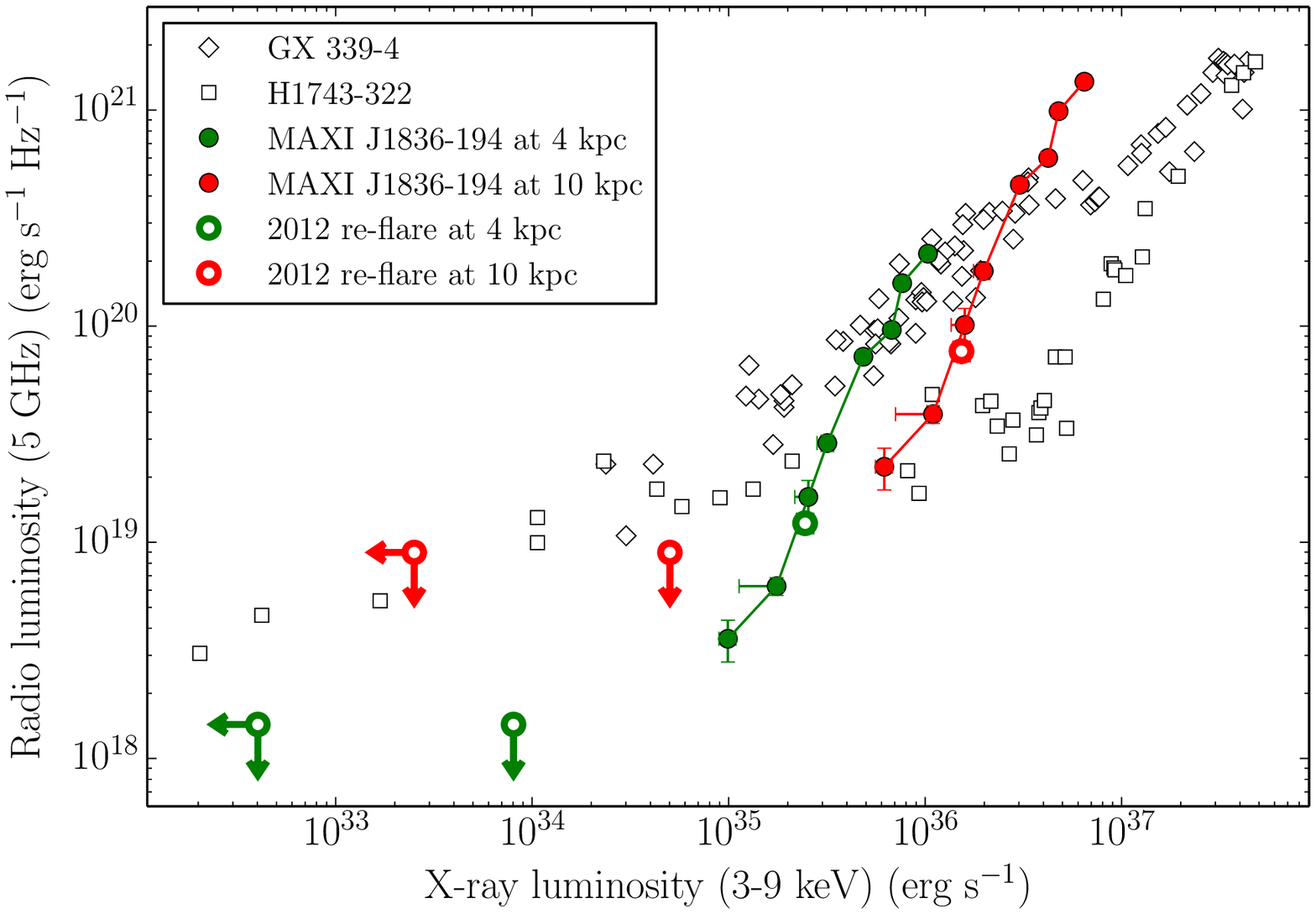}
\caption{The hard-state radio/X-ray correlation of \source{}. The white triangles and squares are representative data from GX 339$-$4 \citep{2013MNRAS.428.2500C} and H1743$-$322 \citep{2011MNRAS.414..677C}, respectively. \source{} data from the 2011 outburst decay are shown by the filled green and red circles for the allowable distance limits of 4 and 10 kpc, respectively, while the open red and green circles show the monitoring from its minor re-brightening in 2012. The observed radio/X-ray correlation for this system is significantly steeper than either of the standard tracks shown by the representative data.}
\label{fig:lrlx_representative}
\end{figure}

\begin{figure}
\centering
\includegraphics[width=\columnwidth]{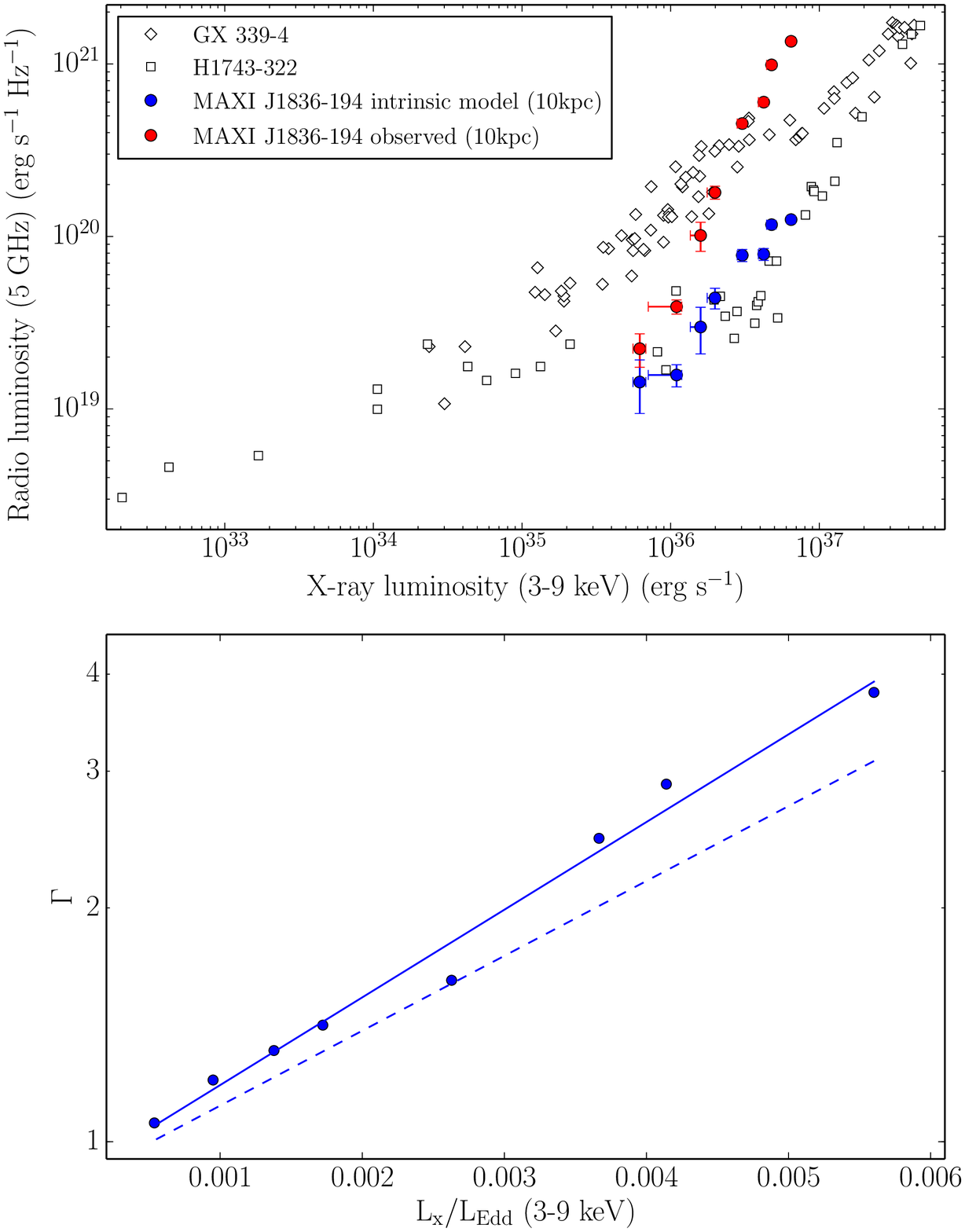}
\caption{Variable boosting of $L_{\rm R}$. Top panel: the hard-state radio/X-ray correlation of \source{} (red points), with the de-boosted, modelled intrinsic data (blue points). Lower panel: $\Gamma$ required to shift $L_{\rm R, intrinsic}$ to $L_{\rm R, observed}$, as a function of X-ray luminosity assuming a distance of 10\,kpc and an inclination of 10$^{\circ}$ \citep{2014MNRAS.439.1381R}. Here, $\log(\Gamma) \propto \kappa l_{\rm x}$, where $l_{\rm x}=L_{\rm x}/{\rm L}_{\rm Edd}$ and $\kappa \approx$ 100--120 (depending on the normalisation). Here, the solid blue and dashed blue lines show results from the minimum and maximum normalisations used to account for the observed scatter about the lower track. The results show that variable Doppler boosting of the radio emission could account for the unusually steep radio/X-ray correlation in this system.}
\label{fig:lrlx}
\end{figure}

The decay phase of the outburst allows us to probe the hard-state radio/X-ray correlation of \source{}. In Figure~\ref{fig:lrlx_representative} we plot the observed hard state radio/X-ray correlation of \source{} with representative data from the upper branch (GX 339$-$4; \citealt{2013MNRAS.428.2500C}) and the lower branch (H1743$-$322; \citealt{2011MNRAS.414..677C}). For comparison, we calculate the luminosities of \source{} over the same energies as \citet{2013MNRAS.428.2500C} and \citet{2011MNRAS.414..677C}. During the decay, we find a steeper than usual radio/X-ray correlation, with an index of 1.8~$\pm$~0.2. This correlation index does not appear to be dependent on frequency (e.g. taking any band of our radio data produces a similar index) and therefore cannot be explained by the changing radio spectral index. Comparing this result with expected relations from \citet{2012MNRAS.419..267P} also shows that the steep correlation cannot be explained by models where the X-ray emission originates either as inverse Compton from a corona, or is dominated by optically thin synchrotron radiation from the jet. We note that our 2011 X-ray observations only span $\sim$1~order of magnitude in X-ray luminosity ($L_{\rm x}$) and deviations from the expected relationship have been found when the X-ray luminosity range is roughly one order of magnitude (thought to be due to changes in the spectral index, electron cooling, additional X-ray emitting regions, etc; \citealt{2013MNRAS.428.2500C}). However, if we include ATCA and {\it Swift} monitoring of the source during (and following) a re-brightening phase in 2012 we extend the X-ray luminosity range. Our 2012 data show that the source appears to lie on same steep track and at 5$\times$10$^{34}$\,erg\,s$^{-1}$ the source is still radio under-luminous compared to other X-ray binaries at comparable $L_{\rm X}$ (Figure~\ref{fig:lrlx_representative}). The relation we observe also holds over $\sim$2.5~orders of magnitude in radio luminosity and does not appear to be random scatter about the standard correlation. Given the low inclination of this system (between 4$^\circ$ and 15$^\circ$; \citealt{2014MNRAS.439.1381R}) it is plausible that variable Doppler boosting affects the correlation. We now investigate this possibility.

Assuming random inclination angles and highly variable boosting, \citet{2011MNRAS.413.2269S} tested a model to determine if variable boosting could account for the scatter of the radio/X-ray correlation for the available sample of black holes. With their model, they were able to qualitatively reproduce both the scatter of the correlation, as well as the `radio-quiet' lower branch but noted that the model had its limitations. Here, we use a similar approach to test if the steep radio/X-ray correlation could be explained by variable relativistic beaming. While recent work has cast doubt over the universality of the two well-defined tracks for the radio/X-ray correlation \citep{2014MNRAS.445..290G}, we assume that the intrinsic radio luminosity of \source{} lies on the lower canonical hard-state tracks. We then determine what boosting would be required to account for the observed radio luminosity, testing whether a variable $\Gamma$ can plausibly account for the steeper than usual correlation. For the system to lie on the upper track, the radio emission at low X-ray luminosities would need to be significantly deboosted (while being boosted at higher X-ray luminosities; Figure~\ref{fig:lrlx_representative}). Deboosted radio emission has been observed in LMXBs (e.g. XTE J1752$-$223; \citealt{2010MNRAS.409L..64Y}) but is unlikely to have affected our observations due to the low inclination angle of the system.

We assume that the X-ray emission remains unbeamed (but see \citealt{2005ApJ...635.1203M,2010MNRAS.405.1759R}) and relate the observed beamed radio luminosity, $L_{\rm R,o}$, to the intrinsic radio luminosity, $L_{\rm R,i}$, by
\begin{equation}
L_{\rm R,o} = L_{\rm R,i}\delta^{n-\alpha}, 
\label{eq:LRoLRi}
\end{equation}
where $\delta$ is the Doppler factor, {\it n}=2 (for a continuously replenished jet; \citealt{2006csxs.book..381F}), and $\alpha$ is the spectral index of the radio emission. The Doppler factor is a function of $\Gamma$ where $\delta=\Gamma^{-1}(1-\beta \cos \theta)^{-1}$, $\beta$ is the velocity of the jet emission as a fraction of the speed of light, and $\theta$ is the inclination of the system; because $\Gamma = (1-\beta^2)^{-1/2}$, we are then able to calculate the Lorentz factor required to shift the modelled intrinsic radio luminosity to that observed.

If we assume an inclination angle of 10$^\circ$ \citep{2014MNRAS.439.1381R}, a source distance of 10 kpc \citep{2014MNRAS.439.1390R} and a jet axis perpendicular to the plane of the binary, then for $L_{\rm R,i}$ to lie on the lower track of the radio/X-ray correlation $\Gamma$ must change by a factor of $\sim$3--4 during the outburst (from $\Gamma \approx$1 at the lowest X-ray luminosity to $\Gamma \sim$3--4 at the highest; Figure~\ref{fig:lrlx}). This result is sensitive to the normalisation of the assumed intrinsic radio luminosity, therefore we have tested a range of normalisations that account for the observed scatter about the lower track. The maximum values of $\Gamma$ are higher than previously thought for both a compact jet (where $\Gamma \le$2; \citealt{2004MNRAS.355.1105F}) and the transient jet of GRS~1915+105 (where 1.6$\le \Gamma \le$1.9; \citealt{2014arXiv1409.2453R}). However, due to the low inclination of \source{}, the values we estimate are not so high that beaming can be ruled out as an explanation for the steep radio/X-ray correlation. Fitting $\Gamma$ against X-ray luminosity (as a fraction of the Eddington luminosity for a 10\,M$_{\odot}$ black hole) gives $\log(\Gamma) \propto \kappa l_{\rm x}$, where $l_{\rm x}=L_{\rm x}/{\rm L}_{\rm Edd}$ and $\kappa \approx$ 100--120. With this result, the bulk Lorentz factor becomes larger than $\sim$1 above 0.05 percent $L_{\rm Edd}$. For comparison, \citet{2011MNRAS.413.2269S} found $\Gamma$ becomes $\ga1$ above 0.1 percent $L_{\rm Edd}$. This level of boosting would be expected to have consequences on the observed radio spectrum, which coupled with the radio spectral index may then provide some insight into physical parameters of the jet (such as the opening angle). However, a full modelling analysis would be required to determine the jet parameters which is beyond the scope of this paper.

At a source distance of $<$8 kpc, the radio emission at low X-ray luminosities falls well below the canonical relation (Figure~\ref{fig:lrlx_representative}) and could not be explained unless it was deboosted. Due to the low inclination angle of the system, we do not expect deboosting. Therefore, if we assume that the steep radio/X-ray correlation is due to a variable Lorentz factor, \source{} must be at a distance greater than 8\,kpc. 

Our results show that relativistic beaming can account for the observed radio/X-ray correlation in \source{}. However, similar to the results of \citet{2011MNRAS.413.2269S}, if we applied our boosting argument to the full sample of LMXBs we would see the correlation break down at X-ray luminosities $\ga0.5$ percent Eddington. Following an outburst, black hole LMXBs generally transition back to the hard state at luminosities between 0.3 and 3~percent Eddington \citep{2013ApJ...779...95K}, with a mean value of $\sim$2 percent \citep{2003A&A...409..697M}, demonstrating that X-ray luminosities of $\ga0.5$ percent Eddington are expected during the hard state decay. \citet{2004MNRAS.355L...1H} analysed the radio/X-ray relation of GX~339$-$4 and V404~Cyg and argued that similar Lorentz factors are expected between sources. We therefore conclude that while a variable Lorentz factor can explain the unusually steep radio/X-ray correlation in \source{}, it cannot be universally applied to the broad sample of black hole binaries. 

It is possible that the steep correlation is a temporary deviation from the standard track \citep{2013MNRAS.428.2500C}. However, when we include data from the 2012 re-flare, we see the source appears to lie on the same steep radio/X-ray track and the steep correlation and does not immediately rejoin the standard track. This system could simply produce an intrinsically steep correlation ($\sim$2$\sigma$ from the steeper track of \citealt{2011MNRAS.414..677C}, or 4$\sigma$ from that of \citealt{2012MNRAS.423..590G}, but see recent results from \citealt{2014MNRAS.445..290G}), however, this index cannot be explained by models of X-ray emission originating as inverse Compton from a corona, or being dominated by optically-thin synchrotron emission from the compact jet.


\section{Conclusions}

We have presented the full VLA, ATCA and VLBA monitoring of \source{} during its 2011 outburst. With our intensive radio coverage of the system we have observed the evolution of the polarised compact jet. We find that the compact jet was linearly polarised at a few percent during the early, bright phase of the outburst. At later times we were unable to place any constraint on the linear polarisation from the source. We determined the VLBI jet axis and found an alignment with the {\it EVPA} (which was calculated from the measured Stokes parameters). This implies that the magnetic field was perpendicular to the compact jet axis, possibly due to a helical structure from the rotation of the accretion disk or compression due to internal shocks within the compact jet. If this alignment is common for all LMXBs then high resolution VLBI observations may not be required to determine the orientation of the jet. However, further studies are required to gauge whether the {\it EVPA} is always aligned with the jet axis. Calculating the peculiar velocity from our astrometric observations, we find that the black hole likely required an asymmetric natal kick during formation to account for its high peculiar velocity. During the decay phase of the outburst we observed an unusually steep radio/X-ray correlation where $L_{\rm R} \propto L_{\rm x}^{1.8\pm0.2}$, much steeper than other black hole LMXB systems. We test whether the steep correlation may be due to a variable jet Lorentz factor. If we assume that the correlation agrees with the radio quiet radio/X-ray correlation track, the Lorentz factor needs to vary from $\Gamma\sim $1 at low X-ray luminosities to $\Gamma \sim $3--4 at high X-ray luminosities to produce the observed correlation. However, this relation is unlikely to explain to the full sample of black hole LMXBs, and it is possible that we are observing a temporary deviation to the standard correlation. Interestingly, data from a re-flare event in 2012 show that the source appears to once again lie on the unusually steep track followed during the 2011 outburst. If the steep radio/X-ray correlation is due only to a variable Lorentz factor, our data require that the source lies at a distance greater than 8\,kpc.

\section*{Acknowledgments}
We would like to thank the anonymous referee for their helpful comments and suggestions. We also thank Tom Maccarone for useful discussions. This research was supported under the Australian Research Council's Discovery Projects funding scheme (project number DP 120102393). DA acknowledges support from the Royal Society. SC acknowledges funding support from the French Research National Agency: CHAOS project ANR-12-BS05-0009 (http://www.chaos-project.fr) and financial support from the UnivEarthS Labex program of Sorbonne Paris Cit\'e (ANR-10-LABX-0023 and ANR-11-IDEX-0005-02). GRS and AJT are supported by an NSERC Discovery Grant. TMB acknowledges support from INAF-PRIN 2012-6. SM acknowledges support by the Spanish Ministerio de Econom\'ia y Competitividad and European Social Funds through a Ram\'on y Cajal Fellowship and the Spanish Ministerio de Ciencia e Innovaci\'on (SM; grant AYA2013-47447-C03-1-P). This research has made use of NASA's Astrophysics Data System. The International Centre for Radio Astronomy Research is a joint venture between Curtin University and the University of Western Australia, funded by the state government of Western Australia and the joint venture partners. The National Radio Astronomy Observatory is a facility of the National Science Foundation operated under cooperative agreement by Associated Universities, Inc. The Australia Telescope Compact Array is part of the Australia Telescope National Facility which is funded by the Commonwealth of Australia for operation as a National Facility managed by CSIRO.

\label{lastpage}


\clearpage

\setcounter{table}{0}
\onecolumn
\centering

\begin{longtable}[c]{ccccccc}
 \caption{{\it continued} VLA flux densities of MAXI~J1836$-$194. $\alpha$ is the radio spectral index. Quoted 1$\sigma$ errors are uncertainties to the fitted source model.}\\

 \hline
Date & MJD & Frequency & Flux density & $Q$               & $U$               & $\alpha$\\
(UT) &     & (GHz)     & (mJy)        & (mJy beam$^{-1}$) & (mJy beam$^{-1}$) &          \\
 \hline
 \endfirsthead
 
 \hline
Date & MJD & Frequency & Flux density & $Q$               & $U$               & $\alpha$\\
(UT) &     & (GHz)     & (mJy)        & (mJy beam$^{-1}$) & (mJy beam$^{-1}$) &          \\
 \hline
 \endhead
 
 \hline
 \endfoot
 
 \hline
 \endlastfoot
2012 Sep 03 & 55807.1 & 4.60  & 27.2$\pm$0.3& 0.24$\pm$0.05 & -0.29$\pm$0.06& 0.64$\pm$0.03 \\
 &  & 7.90  & 38.4$\pm$0.5& 0.67$\pm$0.06 & -0.49$\pm$0.07& \\
2011 Sep 05 & 55809.1 & 5.00  & 23.2$\pm$0.3& 0.08$\pm$0.02 & -0.08$\pm$0.02& 0.84$\pm$0.04\\
 &  & 7.45  & 32.5$\pm$0.4& 0.18$\pm$0.02 & -0.16$\pm$0.02& \\
2011 Sep 07 & 55811.2 & 1.50  & 18.1$\pm$0.4& -- & --& 0.51$\pm$0.01 \\
& & 5.00  & 28.6$\pm$0.7& 0.25$\pm$0.07 & -0.13$\pm$0.07& \\
& & 7.45  & 39.3$\pm$0.9& 0.58$\pm$0.09 & -0.08$\pm$0.08& \\
& & 20.80  & 68$\pm$2& -- & --& \\
& & 25.90  & 79$\pm$2& -- & --& \\
& & 32.02  & 88$\pm$5& -- & --& \\
& & 41.00  & 87$\pm$8& -- & --& \\
2011 Sep 09 & 55813.2 & 1.60  & 24.9$\pm$0.5& -- & --& 0.43$\pm$0.01\\
& & 5.00  & 33.7$\pm$0.6& -0.00$\pm$0.06 & -0.15$\pm$0.05& \\
& & 7.45  & 41.4$\pm$0.9& 0.29$\pm$0.05 & -0.49$\pm$0.05& \\
& & 20.80  & 66$\pm$2& -- & --& \\
& & 25.90  & 73$\pm$2& -- & --& \\
& & 32.02  & 95$\pm$5& -- & --& \\
& & 41.00  & 108$\pm$6& -- & --& \\
2011 Sep 12 & 55816.1 & 1.60  & 36.7$\pm$0.6& -- & --& 0.10$\pm$0.01\\
& & 5.00  & 36.6$\pm$0.8& 0.93$\pm$0.05 & -0.94$\pm$0.05& \\
& & 7.45  & 39.5$\pm$0.8& 0.72$\pm$0.05 & -0.74$\pm$0.04& \\
& & 20.80  & 47$\pm$2& -- & --& \\
& & 25.90  & 48$\pm$2& -- & --& \\
& & 32.02  & 51$\pm$3& -- & --& \\
& & 41.00  & 55$\pm$3& -- & --& \\
2011 Sep 13 & 55817.0 & 1.60  & 29.0$\pm$0.5& -- & --& 0.14$\pm$0.01\\
& & 5.00  & 31.0$\pm$0.4& 0.72$\pm$0.06 & -0.92$\pm$0.06& \\
& & 7.45  & 33.1$\pm$0.4& 0.62$\pm$0.05 & -0.86$\pm$0.05& \\
& & 20.80  & 40.7$\pm$1& -- & --& \\
& & 25.90  & 44$\pm$2& -- & --& \\
& & 32.02  & 46$\pm$3& -- & --& \\
& & 41.00  & 46$\pm$3& -- & --& \\
2011 Sep 18 & 55822.0 & 1.50  & 37$\pm$1& -- & --& 0.13$\pm$0.01\\
& & 5.00  & 34.7$\pm$0.4& 0.85$\pm$0.06 & -0.85$\pm$0.06& \\
& & 7.45  & 36.9$\pm$0.9& 0.62$\pm$0.06 & -0.59$\pm$0.06& \\
& & 20.80  & 44$\pm$1& -- & --& \\
& & 25.90  & 46$\pm$2& -- & --& \\
& & 32.02  & 50$\pm$2& -- & --& \\
& & 41.00  & 441$\pm$3& -- & --& \\
2011 Sep 18 & 55822.9 & 1.56  & 33$\pm$2& -- & --& 0.14$\pm$0.01\\
& & 5.00  & 29.7$\pm$0.4& 0.72$\pm$0.07 & -0.75$\pm$0.07& \\
& & 7.45  & 31.3$\pm$0.4& 0.58$\pm$0.06 & -0.56$\pm$0.06& \\
& & 20.80  & 35.5$\pm$0.7& -- & --& \\
& & 25.90  & 38$\pm$2& -- & --& \\
& & 32.02  & 44$\pm$2& -- & --& \\
& & 41.00  & 37$\pm$2& -- & --& \\
2011 Sep 23 & 55827.061 & 5.00  & 32.7$\pm$0.8& -- & --& 0.34$\pm$0.02\\
& & 7.45  & 34.0$\pm$0.8& -- & --& \\
& & 20.80  & 51$\pm$2& -- & --& \\
& & 25.90  & 54$\pm$2& -- & --& \\
& & 32.02  & 62$\pm$3& -- & --& \\
& & 41.00  & 66$\pm$4& -- & --& \\
2011 Sep 26 & 55830.9 & 5.00  & 14.1$\pm$0.4& -- & --& 0.51$\pm$0.02\\ 
& & 7.45  & 14.8$\pm$0.4& -- & --& \\
2011 Sep 26 & 55830.9 & 20.80  & 27$\pm$1& -- & --& \\
& & 25.90  & 30$\pm$1& -- & --& 0.51$\pm$0.02\\
& & 31.52  & 36$\pm$2& -- & --& \\
& & 37.52  & 40$\pm$2& -- & --& \\
2011 Sep 30 & 55834.0 & 5.26  & 11.3$\pm$0.3& -- & --& 0.57$\pm$0.02\\
& & 7.45  & 11.7$\pm$0.3& -- & --& \\
& & 20.80  & 23.5$\pm$0.9& -- & --& \\
& & 25.90  & 26$\pm$1& -- & --& \\
& & 31.52  & 30$\pm$2& -- & --& \\
& & 37.52  & 32$\pm$2& -- & --& \\
2011 Oct 07 & 55841.9 & 5.26  & 8$\pm$0.3& -- & --& 0.59$\pm$0.03\\
& & 7.45  & 8$\pm$0.3& -- & --& \\
& & 20.80  & 17.5$\pm$0.9& -- & --& \\
& & 25.90  & 19.3$\pm$0.9& -- & --& \\
& & 31.52  & 22$\pm$2& -- & --& \\
& & 37.52  & 25$\pm$2& -- & --& \\
2011 Oct 12 & 55846.0 & 5.26  & 5.0$\pm$0.3& -- & --& 0.55$\pm$0.03\\
& & 7.45  & 5.7$\pm$0.2& -- & --& \\
& & 20.80  & 9.8$\pm$0.5& -- & --& \\
& & 25.90  & 12.2$\pm$0.7& -- & --& \\
& & 31.52  & 14$\pm$1& -- & --& \\
& & 37.52  & 14$\pm$1& -- & --& \\
2011 Oct 22 & 55856.9 & 5.26  & 3.8$\pm$0.2& -- & --& 0.25$\pm$0.04\\
& & 7.45  & 4.0$\pm$0.2& -- & --& \\
& & 20.80  & 5.3$\pm$0.4& -- & --& \\
& & 25.90  & 5.6$\pm$0.4& -- & --& \\
& & 31.52  & 5.8$\pm$0.6& -- & --& \\
& & 37.52  & 5.7$\pm$0.8& -- & --& \\ 
2011 Nov 01&55866.9 & 5.26  & 1.5$\pm$0.1& -- & --& 0.33$\pm$0.06\\
& & 7.45  & 1.46$\pm$0.08& -- & --& \\
& & 20.80  & 1.9$\pm$0.2& -- & --& \\
& & 25.90  & 2.6$\pm$0.4& -- & --& \\
& & 31.52  & 2.6$\pm$0.4& -- & --& \\
& & 37.52  & 3.2$\pm$0.7& -- & --& \\
2011 Nov 11 & 55876.8 & 5.26  & 0.8$\pm$0.2& -- & --& 0.4$\pm$0.1\\
& & 7.45  & 0.8$\pm$0.1& -- & --& \\
& & 20.80  & 1.0$\pm$0.1& -- & --& \\
& & 25.90  & 1.5$\pm$0.2& -- & --& \\
2011 Nov 18 & 55883.9 & 5.26  & 0.33$\pm$0.03& -- & --& 0.5$\pm$0.1\\
& & 7.45  & 0.31$\pm$0.06& -- & --& \\
& & 20.80  & 0.6$\pm$0.1& -- & --& \\
& & 25.90  & 0.8$\pm$0.1& -- & --& \\
2011 Dec 03 & 55898.8 & 5.26  & 0.19$\pm$0.04& -- & --& 0.7$\pm$0.1\\
 &  & 7.45  & 0.25$\pm$0.02& -- & --& \\
 &  & 20.80  & 0.49$\pm$0.07& -- & --& \\
 &  & 25.90  & 0.5$\pm$0.1& -- & --& \\
                                        
\hline
\end{longtable}

\end{document}